\input amstex
\magnification=1200
\documentstyle{amsppt}

\widestnumber\no{999}
\TagsOnRight
\NoRunningHeads
\NoBlackBoxes

\baselineskip 20pt

\hsize 6.25truein
\vsize 8.75truein

\catcode`\@=11
\def\logo@{}
\catcode`\@=13

\def\tr{\text{\rm tr}}

\def\det{\text{\rm det}}

\topmatter
\title Regularity of the Einstein Equations at Future Null Infinity \endtitle
\author Vincent Moncrief $^{1}$  and Oliver Rinne $^{2,3,4}$ \endauthor 
\affil $^1$ Department of Mathematics and \\ Department of Physics \\
Yale University \\ New Haven, CT 06520, USA \\ 
{\tt vincent.moncrief{\@}yale.edu} \\ \\
$^2$ Department of Applied Mathematics \\ and Theoretical Physics  \\
Centre for Mathematical Sciences, Wilberforce Road \\ Cambridge CB3
0WA, UK\\ {\tt o.rinne{\@}damtp.cam.ac.uk}   \\ \\
$^3$  King's College, Cambridge CB2 1ST, UK \\ \\
$^4$  Theoretical Astrophysics 130-33\\ California Institute of Technology\\
1200 East California Boulevard\\ Pasadena, CA 91125, USA \\  \endaffil

\abstract When Einstein's equations for an asymptotically flat, vacuum spacetime are reexpressed in terms of an appropriate conformal metric that is regular at (future) null infinity, they develop apparently singular terms in the associated conformal factor and thus appear to be ill-behaved at this (exterior) boundary.  In this article however we show, through an enforcement of the Hamiltonian and momentum constraints to the needed order in a Taylor expansion, that these apparently singular terms are not only regular at the boundary but can in fact be explicitly evaluated there in terms of conformally regular geometric data.  Though we employ a rather rigidly constrained and gauge fixed formulation of the field equations, we discuss the extent to which we expect our results to have a more `universal' significance and, in particular, to be applicable, after minor modifications, to alternative formulations. \endabstract
\endtopmatter

\head I. Introduction\endhead
\vskip .10in

The natural outer boundary for gravitational radiation problems in asymptotically flat spacetimes is future null infinity (also referred to as  $\Cal I^+$ or `Scri') \cite{1,2,3,4}. 
The feasibility of using hyperboloidal slicings of spacetime and putting the outer boundary of numerical calculations at Scri has long been recognized, thanks to the elegant work of H. Friedrich and his associates who have numerically implemented Friedrich's conformally regular field equations for this purpose \cite{5,6}.  In spite of this success, most numerical relativists attempting to solve the black hole collisions problem prefer to use some direct formulation of the Einstein equations themselves (which, though implied by Friedrich's conformally regular system, are not explicitly included within it).  Since, however, Einstein's equations are not conformally regular in any obvious sense, it has heretofore not seemed feasible to put the outer boundary of such, more conventional numerical calculations at Scri.

In this paper however, we propose a specific (constrained and gauge fixed) formulation of the Einstein equations for which it does seem feasible to put the exterior boundary at future null infinity.  In particular, we show that when the usual Hamiltonian and momentum constraints are enforced (at least to a suitable order in a Taylor expansion about Scri) and when the well known geometrical condition that Scri should be shear-free is taken into account, then all of the apparently singular terms in the Einstein evolution equations (that appear when the latter are reexpressed in terms of the conformal metric) are not only actually regular at this boundary, but also can in fact be explicitly evaluated.  We further show that the conditions at Scri that were needed to establish this regularity (i.e., the vanishing of the shear and satisfaction of the constraints to the requisite order) are preserved by the boundary forms of the evolution equations so derived.

Our particular formulation of the field equations was motivated, in part, by the proof, given in Ref. \cite{7}, that the usual vacuum Einstein equations expressed in CMCSH (constant-mean-curvature-spatial-harmonic) gauge satisfy a well-posed Cauchy evolution theorem in the `cosmological' setting of spatially compact spacetimes.  The constant mean curvature gauge condition serves to partially decouple the constraints and to define a convenient slicing for spacetime whereas the spatial harmonic condition effectively reduces the evolution equations for the spatial metric to hyperbolic form.  The remaining spacetime metric components (i.e., the lapse function and shift vector field) are determined by elliptic equations that enforce the continuation of the gauge conditions.  This theorem, on the other hand, does not treat the decomposition of the spacetime metric into the product of a conformal metric and a conformal factor nor does it address constructs appropriate to asymptotically flat spacetimes such as the attachment of a conformal boundary.  Thus it does not literally apply to the formulation proposed here.  Nevertheless its existence suggested the feasibility of adopting analogous gauge conditions (wherein the physical mean curvature is held constant but the {\it conformal} metric is instead subjected to the spatial harmonic condition) for the present problem.

In spite of the built-in rigidity of our particular setup, we are confident that the central results of the present paper have a more `universal' significance and do not require, for their application, a strict adherence to our specific formulation of the field equations.  Rather we believe that our main results can be applied to a wide variety of alternative formulations that employ, for example, different gauge conditions or allow free (i.e., unconstrained) evolutions provided suitable attention is paid to the necessary regularity conditions at Scri. We shall return to and clarify these remarks about `universality' later in the presentation.

Though a full numerical implementation of our ideas has not yet been 
carried out, one of us (O.R.) is planning
to modify his existing vacuum, axisymmetric evolution code \cite{8} so as to fix the outer boundary at Scri using a variant of the ideas described herein. Furthermore, L. Buchman and H. Pfeiffer have established the feasibility of numerically solving the constraint equations out to Scri by explicitly computing initial data sets for multiple, boosted spinning black holes \cite{9}. In addition the linearization of our proposed formulation, about a conformally compactified Schwarzschild background, has been carried out and partially analyzed with a view towards developing code tests for the nonlinear calculations \cite{10}.

\head II. Constraint and Evolution Equations\endhead
\vskip .10in

Our constrained evolution scheme involves solving elliptic equations for the conformal factor, the conformal lapse and physical shift and for the unphysical (i.e., conformal) mean curvature. The conformal spatial metric and a momentum tensor density (the trace-free part of the physical ADM momentum) are evolved. The constant time slices are CMC slices which extend outward to future null infinity which is fixed to coincide with a coordinate cylinder in the conformal spacetime metric through the imposition of suitable boundary conditions upon the conformal factor, the conformal lapse and the physical shift.

The physical ADM variables are $\{ g_{ij}, \pi^{ij}\}$ and $\{ N, X^i\}$, where $g_{ij}$ is the spatial metric, $\pi^{ij}$ is a momentum tensor density (i.e., $\pi^{ij}/\mu_g$, with $\mu_g = \sqrt{\det ~ g_{mn}}$, is a tensor) defined in terms of the extrinsic curvature $K_{mn}$ via
$$
\pi^{ij} = \mu_g (g^{ij} g^{mn} - g^{im} g^{jn})K_{mn}, \tag2.1
$$
$N$ is the lapse and $X^i$ is the shift. 
In our conventions $K_{mn}$ is defined via 
$g_{mn,t} = - 2 N K_{mn} + X^i g_{mn,i} + X^i_{,m} g_{in} + X^i_{,n} g_{mi}$.
We use Latin (Greek) indices to denote spatial (spacetime) indices.
The spacetime metric has the line element
$$
ds^2 = - N^2 dt^2 + g_{ij} (dx^i + X^idt)(dx^j + X^jdt) 
 = {}^{(4)} g_{\mu\nu} dx^\mu dx^\nu \tag2.2
$$
and we write this alternatively as
$$
ds^2 = \frac{ \{-\tilde N^2 dt^2+\gamma_{ij}(dx^i+X^idt)(dx^j
  +X^jdt)\}}{\Omega^2}  = \frac{{}^{(4)} \gamma_{\mu\nu} dx^\mu
  dx^\nu}{\Omega^2}, \tag2.3
$$
where $\Omega$ is the conformal factor, $\gamma_{ij}$ the conformal
spatial  metric and $\tilde N$ the conformal lapse.  Notice that $X^i$
serves as both physical and conformal shift but that $g_{ij},
\gamma_{ij}, N$ and $\tilde N$ are related by
$$
g_{ij} = \frac{1}{\Omega^{2}}~ ~ \gamma_{ij}, ~ ~ N = \frac{\tilde N}{\Omega}. \tag2.4
$$
$\Omega$ is greater than zero in the interior but approaches zero at $\Cal I^+$.

CMC slicing is defined by
$$
\frac{\tr_{g}\pi}{\mu_{g}} = - 2K = ~ \text{constant} \tag2.5
$$
where $\tr_g\pi = g_{ij}\pi^{ij}$ and $K = -g_{ij} K^{ij}= - (\text{mean ~ curvature})$.
The slightly odd sign convention chosen for the definition of $K$ has been made so that this constant will be positive.  The traceless part, $\pi^{\tr\,{ij}}$, of the ADM momentum is defined by
$$
\pi^{\tr\,{ij}} := \pi^{ij} - \frac{1}{3} g^{ij} g_{mn} \pi^{mn}. \tag2.6
$$

When $K$ is a constant the momentum constraint can be written
$$
\tilde\nabla_j \left[\frac{\pi^{\tr\,{ij}}}{\Omega^{2}}\right] = 0 \tag2.7
$$
where $\tilde\nabla_i$ is the covariant derivative with respect to $\gamma_{ij}$.  Defining, as usual, the mixed physical components of $\pi^{\tr}$ via $\pi^{\tr\,{j}}_i = g_{im} \pi^{\tr\,{mj}}$ the momentum constraint can also be written as
$$
\tilde\nabla_j\pi^{\tr\,{j}}_i = 0; ~ ~ \pi^{\tr\,{i}}_i = 0. \tag2.8
$$
Standard methods are available for solving this equation through the use of orthogonal decompositions.

The Hamiltonian constraint (also known as the Lichn\'erowitz equation) has the form
$$\align
&-4 \Omega(\tilde\nabla^j \tilde\nabla_j\Omega) + 6\gamma^{ij} \Omega_{,i}\Omega_{,j} - \tilde R(\gamma)\Omega^2 \tag2.9 \\
&-\frac{2}{3} K^2 + \frac{\pi^{\tr\,{ij}}\pi^{\tr\,{kl}}}{(\mu_\gamma)^{2}} \gamma_{ik}\gamma_{jl} \Omega^2 = 0,
\endalign
$$
where $\tilde\nabla^j \tilde\nabla_j = \gamma^{ij} \tilde\nabla_i\tilde\nabla_j$ is the Laplacian with respect to $\gamma_{ij}, \tilde R(\gamma)$ is the scalar curvature of $\gamma_{ij}$ and $\mu_\gamma = \sqrt{\det ~\gamma_{mn}}$.  This elliptic equation for $\Omega$ degenerates at $\Cal I^+$ where the conformal factor vanishes.

From the ADM evolution equations, one can easily derive (by taking the trace of the $g_{ij,t}$ equation) the following evolution equation for $\Omega$, 
$$
6 \left[\Omega,_t - X^i \, \Omega,_i \right]-\Omega \, \Gamma = -2 \,
K \tilde{N}, \tag2.10
$$
where $\Gamma$ is defined by
$$
\Gamma := \gamma^{mn} \gamma_{mn,t} - 2 \, \tilde{\nabla}_l X^l. \tag2.11
$$
To fix the decomposition of $g_{ij}$ into a conformal metric and a
conformal factor, we need to impose a normalization condition upon
$\gamma_{ij}$.  A mathematically appealing choice is to exploit
Yamabe's theorem \cite{11}
and demand that $\tilde R(\gamma)$ be a (spacetime) constant.  The corresponding requirement that $\partial_t \tilde R(\gamma) = 0$ then leads to an elliptic equation for $\Gamma$ given by
$$
\frac{2}{3} \, \gamma^{ij} \left(\tilde{\nabla}_i \tilde{\nabla}_j
\Gamma \right) +\frac{1}{3} \, \tilde{R}(\gamma) \, \Gamma =
\tilde{\nabla}_i \tilde{\nabla}_j \left[\frac{2
\tilde{N}}{\mu_{\gamma}}~ \pi^{\tr\,ij}\right]
-\tilde{R}_{ij}(\gamma)\left[\frac{2 \tilde{N}}{\mu_{\gamma}}~
\pi^{\tr\,ij}\right], \tag2.12
$$
where $\tilde R_{ij}(\gamma)$ is the Ricci tensor of $\gamma_{ij}$ and, as defined above, $\tilde N$ is the conformal lapse function $(\tilde N := N\Omega)$.  Note also, from Eq. (2.11), that the quantity $-\Gamma/(2\tilde N)$ is the unphysical mean curvature.

$\tilde N$ is determined by solving the elliptic equation
$$\align
0 &= - \, \Omega^2 \left(\gamma^{ij} \, \tilde{\nabla}_i
\tilde{\nabla}_j \tilde{N} \right) + 3 \, \Omega \, \gamma^{ij}
\tilde{N},_i \, \Omega,_j - \frac{3}{2} \, \tilde{N} \, \gamma^{ij}
\Omega,_i \, \Omega,_j + \frac{\tilde{N}}{6} \, K^2  \tag2.13 \\ 
&- \frac{\tilde{N}\Omega^2}{4} \tilde{R}(\gamma) + \frac{5}{4} \,
\frac{\tilde{N}}{{\mu_\gamma}^2} \; \Omega^2 \, \gamma_{il} \,
\gamma_{jm} \, \pi^{\tr\,im} \, \pi^{\tr\,jl}
\endalign
$$
which enforces the condition $\partial_t K = 0$ (assuming that $K$ is spatially constant).  By combining Eqs. (2.9) and (2.13) in a straightforward way, one can derive a less degenerate form for the conformal lapse equation which has only a single power of $\Omega$ multiplying the Laplacian of $\tilde N$.

For the physical shift vector $X^i$, we propose to determine it so as to preserve the spatial harmonic gauge condition defined by
$$
V^k := \gamma^{ij} (\tilde\Gamma^k_{ij} (\gamma) - \tilde\Gamma^k_{ij} (\overset\circ\to\gamma)) = 0 \tag2.14
$$
where $\overset\circ\to\gamma$ is a fixed (i.e., time-independent)
reference metric (for example $\overset\circ\to\gamma_{ij} =
\gamma_{ij}\mid_{t=0}$ is a possible choice).
The Christoffel symbols of $\gamma$ and $\overset\circ\to\gamma$ are
denoted by $\tilde\Gamma^k_{ij}(\gamma)$ and
$\tilde\Gamma^k_{ij}(\overset\circ\to\gamma)$, respectively.
  Equation (2.14) corresponds to the demand that the identity map from $(M, \gamma_{ij})$ to $(M,\overset\circ\to\gamma_{ij})$ be harmonic.  The shift equation results from requiring $\partial_tV^k = 0$ and is given by
$$\align
0 &= -\frac{1}{3}\,\Gamma \,V^k - \frac{1}{6}\, \gamma^{kl} \,
\Gamma,_l ~-~ \frac{2 \, \tilde{N}}{\mu_\gamma} \, \pi^{\tr\,ij} \,
\left(\tilde{\Gamma}_{ij}^k(\gamma)-\tilde\Gamma_{ij}^k(\overset\circ\to\gamma)\right)\tag2.15 \\
&+ \tilde{\nabla}_l~\left[\frac{2 \,
\tilde{N}}{\mu_\gamma} \, \pi^{\tr\,kl} \right] ~+~
\tilde{\nabla}_j~\left[ \tilde{\nabla}^j\,X^k +
\tilde{\nabla}^k\,X^j\right] ~-~ \tilde{\nabla}^k~\left(
\tilde{\nabla}_j\,X^j\right)\\ 
&- \left(\tilde{\nabla}^i\,X^j ~+~ \tilde{\nabla}^j\,X^i\right)~\left(\tilde{\Gamma}_{ij}^k(\gamma)
-\tilde\Gamma_{ij}^k(\overset\circ\to\gamma)\right).
\endalign
$$

An extensive mathematical study of the use of constant-mean-curvature-spatial-harmonic (or CMCSH for brevity) gauge conditions in a `cosmological' (i.e. spatially compact) setting was made by the authors of Ref. [7] who proved a well-posedness theorem for the vacuum Einstein equations (in arbitrary spacetime dimension) in this gauge. Their theorem did not deal with the conformal decomposition of the spacetime metric or with the presence of future null infinity in the conformal geometry and so, strictly speaking, is not applicable to the problem dealt with here. Nevertheless, their formulation provided some of the motivation for our setup and might conceivably provide the model for the well-posedness theorem that one would eventually like to prove for the equations we are studying. It is worth mentioning here that the main reason for the choice of the spatial harmonic gauge condition made in Ref. [7] was the fact that it nullifies terms in the Ricci tensor of $g_{ij}$ that, if present, would disturb the hyperbolic character of the equations of motion for this metric. The strategy is rather similar to that motivating the use of {\it spacetime} harmonic coordinates in other formulations (to nullify corresponding terms in the spacetime Ricci tensor) but leaves open the possibility for determining the lapse and shift through the solution of elliptic equations instead of hyperbolic ones. Of course the resulting evolution system is now hyperbolic/elliptic rather than purely hyperbolic. In our formulation the conformal metric $\gamma_{ij}$ (normalized by the condition $\tilde R(\gamma) =$ constant and gauge fixed by the spatial harmonic conditions $V^k = 0$) represents the two propagating degrees of freedom of the gravitational field.

At $\Cal I^+$, where $\Omega$ vanishes, Eq. (2.9) forces the gradient of $\Omega$ to satisfy
$$
\gamma^{ij} \Omega_{,i}\Omega_{,j}\mid_{\Cal I^{+}} = \left( \frac{K}{3}\right)^2 \tag2.16
$$
whereas Eq. (2.10) yields
$$
X^i \Omega_{,i}\mid_{\Cal I^{+}} = \frac{1}{3} ~ ~ K\tilde N\mid_{\Cal I^{+}}. \tag2.17
$$
The shift field at $\Cal I^+$ must therefore take the form
$$
X^i\mid_{\Cal I^{+}} = \frac{3}{K} ~ \tilde N \gamma^{ij} \Omega_{,j}\mid_{\Cal I^{+}} + Z^i\mid_{\Cal I^{+}} \tag2.18
$$
where the vector field $Z^i\mid_{\Cal I^{+}} \frac{\partial}{\partial x^i}$ is purely tangential to $\Cal I^+$ (i.e., satisfies \newline $Z^i\mid_{\Cal I^{+}} \Omega_{,i}\mid_{\Cal I^{+}} = 0$). Note that the above then implies
$$
\gamma_{ij} X^iX^j\mid_{\Cal I^{+}} = (\tilde N^2 + \gamma_{ij} Z^iZ^j)\mid_{\Cal I^{+}} . \tag2.19
$$
If, however, we wish to have the time coordinate vector field $\frac{\partial}{\partial t}$ tangent to the null generators of $\Cal I^+$ (i.e., to be such that $\frac{\partial}{\partial t}$ is null at $\Cal I^+$ in the conformal geometry) then we would need to have
$$
\gamma_{ij} X^iX^j\mid_{\Cal I^{+}} = \tilde N^2\mid_{\Cal I^{+}} \tag2.20
$$
satisfied there.  In other words, the boundary condition $Z^i\mid_{\Cal I^{+}} = 0$ corresponds to choosing $\frac{\partial}{\partial t}\mid_{\Cal I^{+}}$ to be null, rather than spacelike, at the boundary.  For the `normal' component of $X^i$ at $\Cal I^+$ however, one has no flexibility since the above choice is forced by the requirement that $\Cal I^+$ coincide with a fixed cylinder in the conformal geometry.

Finally, the evolution equations for $\{\gamma_{ij}, \pi^{\tr\,{ij}}\}$ are given by
$$
\partial_t \gamma_{ij} = \frac{2 \tilde{N}}{\mu_\gamma}\,\gamma_{i
l}\,\gamma_{jm}\, \pi^{\tr\,l m} + \frac{1}{3} \, \gamma_{ij} \,
\Gamma +\gamma_{i l}\,\gamma_{jm}\,\left(\tilde{\nabla}^l X^m
+\tilde{\nabla}^m X^l \right), \tag2.21
$$
and
$$\align
\partial_t \, \pi^{\tr\,ij} &= \left( X^m \, \pi^{\tr\,ij}
\right),_m - X^i,_m \, \pi^{\tr\,mj} - X^j,_m \, \pi^{\tr\,im} -
\frac{2 \tilde{N}}{\mu_\gamma} \, \pi^{\tr\,im} \, \pi^{\tr\,jl}
\, \gamma_{lm}  \\
&\qquad\qquad\qquad - \frac{2}{3} \frac{\tilde{N}}{\Omega} \,
\pi^{\tr\,ij} \, K  \tag2.22 \\
&+ \mu_\gamma \left(\tilde{\nabla}^i \tilde{\nabla}^j \tilde{N} -\frac{1}{3} \,
\gamma^{ij} \, \gamma^{mn} \tilde{\nabla}_m \tilde{\nabla}_n
\tilde{N}\right) - \mu_\gamma \, \tilde{N}
\left(\tilde{R}^{ij}(\gamma) - \frac{1}{3} \, \gamma^{ij} \,
\tilde{R}(\gamma) \right) \\
&-  2 \, \mu_\gamma \, \tilde{N} \left(\frac{\tilde{\nabla}^i \tilde{\nabla}^j
\Omega}{\Omega} - \frac{1}{3} \, \gamma^{ij} \,
\gamma_{mn}\frac{\tilde{\nabla}^m \tilde{\nabla}^n
\Omega}{\Omega}\right).
\endalign
$$
The main content of this paper involves analyzing the apparently singular, $\Omega$-dependent terms in Eq. (2.22).  We shall, in fact, derive explicitly regular forms for the evolution equations at $\Cal I^+$ and show how these can be used to propagate geometric data along this conformal boundary.

In effect, our formulation is maximally elliptic in that we envision solving elliptic equations for the Hamiltonian and momentum constraints, Eqs. (2.9) and (2.7), the conformal lapse and physical shift, Eqs. (2.13) and (2.15) as well as Eq. (2.12) for the unphysical mean curvature.  This could only be numerically practical through the use of elliptic solvers that operate at the level of `linear complexity' but fortunately several such systems are currently available.  However, as we shall emphasize throughout this paper, our main conclusions should hold without the need for a strict adherence to the constrained evolution program outlined above but should instead be applicable, with minor modifications, to a wide variety of alternative schemes.

\head III. Regularity at Future Null Infinity\endhead
\vskip .10in

The constraint and evolution equations presented in section II, as well as the elliptic equation for the conformal lapse function $\tilde N$, are formally singular at Scri where the conformal factor vanishes.  In this section, we analyze the behavior of the corresponding fields in a neighborhood of this boundary and derive a set of regularity conditions that sufficiently differentiable solutions must satisfy at Scri.  Later we shall establish the consistency of the regularity conditions by showing that they are preserved under time evolution.  This last step will necessitate explicit evaluation of the evolution equations, including their apparently singular terms, at the conformal boundary.

Let $M$ be a three-dimensional spacelike slice with conformal boundary $\partial M \approx S^2$ on $\Cal I^+$.  We choose coordinates $\{ x^i\} = \{ (x^1, x^a)\mid a = 2,3\} = \{(r,\theta,\varphi)\}$ for the Riemannian manifold (with boundary) $(M,\gamma_{ij})$ such that $x^1 = r$ is a `radial' coordinate satisfying $r\leq r_+ =$ constant on $M$ with $r = r_+$ corresponding to the boundary $2$-sphere $\partial M$ at Scri.  The indices $a,b,\ldots$ are restricted to range over $2$ and $3$ so that $\{x^a\} = \{(\theta,\varphi)\}$ are `angular' coordinates for the $r =$ constant surfaces.  Aside from requiring that $\partial M$ coincide with the coordinate sphere defined by $r = r_+$, we leave, for the moment, the choice of coordinates on $M$ arbitrary.

We suppose that each of the relevant fields can be expanded in a finite Taylor series (with remainder) about the boundary at $r = r_+$.  Thus for each $u\in \{\Omega, \gamma_{ij}, \pi^{\tr\,{ij}},\tilde N\}$ there is an integer $l > 0$ (depending on the choice of $u$) such that $u$ can be expressed as
$$\align
&u(x^i) = u_0(x^a) + u_1(x^a)(r-r_+) \tag3.1 \\
&+\frac{1}{2!} u_2(x^a) (r-r_+)^2 +\ldots + \frac{1}{l !} u_l (x^a) (r-r_+)^l \\
&+ \text{rm}(x^i),
\endalign
$$
with $u_k := \lim\limits_{r\nearrow r_{+}} (\partial^k_r u)$ and remainder $\text{rm}(x^i) = o((r-r_+)^l)$, on some interior neighborhood of $\partial M$.  In the following, it will be convenient to utilize the symbol 
$\overset{\wedge}\to {=}$ to denote equality at $\partial M$.  By definition the conformal factor thus satisfies 
$\Omega \overset{\wedge}\to {=} 0$ so that, in particular, $\Omega_0 = 0$ in the corresponding Taylor expansion. A remarkable feature of the degenerating elliptic equations that we have to deal with is that they permit one to explicitly compute more detailed asymptotic information (in the form of Taylor expansions) about the corresponding solutions than would be possible in the case of non-degenerate equations. In particular, we shall be able to evaluate the first three radial derivatives of $\Omega$, the first two such derivatives of $\tilde N$ and the first radial derivative of the $\pi^{\tr\,{ri}}$ components of $\pi^{\tr\,{ij}}$ `universally' at Scri (i.e., expressible in terms of data there without reference to the actual global solutions). Remarkably these particular derivatives are precisely what is needed to then evaluate the evolution equations (including their apparently singular terms) at Scri and to verify that they imply the preservation of the associated regularity conditions.

A rigorous treatment of the constraint equations on `hyperboloidal' initial data surfaces intersecting $\Cal I^+$ has already been given by Andersson, Chru\'sciel and Friedrich in an important series of papers from the early 90's \cite{12, 13}. In particular they derived the regularity conditions needed for a differentiable Scri in terms of ADM Cauchy data. Our main contribution here is to carry this analysis a step further and show how one can evaluate the (apparently singular) evolution equations at Scri and use them to verify preservation of the regularity conditions within the framework of our particular gauge fixed evolutionary formulation. Though we use this framework in order to have a complete, coherent system for calculations, we do not believe, as emphasized in the introduction, that our main conclusions hinge crucially upon its specific form but rather that they should apply equally well to a variety of other formulations of the field equations. Though the arguments sketched above yield expressions for the first radial derivative of only the $ri$ components of $\pi^{\tr\,{ij}}$ at Scri we shall present, in section V below, an alternative method for evaluating the apparently singular terms in the $\pi^{\tr\,{ij}}$ evolution equations that will finally allow us to compute the first radial derivatives of the angular components, $\pi^{\tr\,{ab}}$, at Scri as well.

As in section II, we assume that the Riemannian manifold $(M,\gamma_{ij})$ corresponds to a CMC slice in the physical spacetime and take the mean curvature of the latter to be a negative constant (written as before as $-K, K > 0$, constant). It is convenient to reexpress the conformal metric $\gamma_{ij}$ relative to the chosen coordinates $\{x^i\} = \{ (r,\theta,\varphi)\}$ in $2+1$ dimensional (Riemannian) ADM form by setting
$$\align
dl^2 &= \gamma_{ij} dx^i dx^j \tag3.2 \\
&= n^2 dr^2 + h_{ab}(dx^a + Y^a dr)(dx^b + Y^b dr).
\endalign
$$
The induced metric on an $r =$ constant surface thus has the line element
$$
d\sigma^2 = h_{ab} dx^a dx^b\bigm|_{r = \text{constant}} \tag3.3
$$
and we let the symbols $\vert a$ or $^{(2)}\nabla_a(h)$ signify covariant differentiation with respect to this metric.

The unit {\it outward} pointing normal field to an $r =$ constant surface in $(M, \gamma_{ij})$ is given in coordinates by
$$
(\nu^i) = (\frac{1}{n} , \frac{-Y^{a}}{n}) \tag3.4
$$
or, equivalently, in covariant form, by 
$$
(\nu_i) = (n, 0, 0). \tag3.5
$$
The second fundamental form $\lambda_{ab}$, induced by $\gamma_{ij}$ on such $r =$ constant surfaces is defined via
$$
\lambda_{ab} = - \tilde\nabla_b\nu_a = - \nu_{a;b} \tag3.6
$$
(where, as before, $\tilde\nabla_i$ or $;i$ signifies covariant differentiation with respect to $\gamma_{ij}$). Written out explicitly Eq. (3.6) leads to
$$
h_{ab,r} =  -2n\lambda_{ab} + Y_{a\mid b} + Y_{b\mid a} \tag3.7
$$
where $Y_a := h_{ab} Y^b$.  Since expressions for the Ricci tensor
components, $\tilde R_{ij}(\gamma)$, of $\gamma_{ij}$ are needed for some of the calculations we give these explicitly, in the present notation, in the appendix below.

The Hamiltonian constraint (Eq. (2.9)) gives, using $\Omega \overset{\wedge}\to {=} 0$, the following equation at $\partial M$:
$$
\{ \gamma^{ij} \Omega_{,i}\Omega_{,j}\}\bigm|_{r=r_{+}} \overset{\wedge}\to {=} (\frac{K}{3})^2 . \tag3.8
$$
Since $K > 0$ and we require $\Omega > 0$ for $r < r_+$ this yields
$$
\frac{\Omega_{,r}}{n}\bigm|_{r=r_{+}} \overset{\wedge}\to {=} - \frac{K}{3}. \tag3.9
$$
One is free to compute angular (but not radial) derivatives of such equations and deduce, for example, formulas such as
$$
(\frac{\Omega_{,r}}{n})_{,a} \overset{\wedge}\to {=} 0, \ldots . \tag3.10
$$

To get the next order Taylor coefficient of $\Omega$, we compute the radial derivative of Eq. (2.9) and evaluate the result at $\partial M$ to find
$$
\left\{\Omega_{,rr} - \Omega_{,r} \left[ \frac{n_{,r}}{n} + \frac{Y^{a}n_{\mid a}}{n} - \frac{n}{2} h^{ab} \lambda_{ab}\right]\right\}\bigm|_{r=r_{+}} \overset{\wedge}\to {=} 0. \tag3.11
$$
The third radial derivative of $\Omega$ at $\partial M$ can be computed by first taking the Laplacian of Eq. (2.9) and then reducing the resulting expression using the foregoing results.  The key formula resulting from this calculation can be written as
$$\align
&\left\{ \frac{1}{n} (\gamma^{ij} \tilde\nabla_i\tilde\nabla_j \Omega)_{,r}\right\} \biggm|_{r=r_{+}} \overset{\wedge}\to {=} \\
&K\biggl\{ \lambda_{ab}\lambda_{cd} h^{ac}h^{bd} - \frac{1}{2} (h^{ab} \lambda_{ab})^2 + n^2 \tilde R^{rr}(\gamma) \\
&- \frac{1}{6} \tilde R(\gamma) + \frac{1}{6} \frac{\pi^{\tr\,{l j}}\pi^{\tr\,{im}}}{(\mu_{\gamma})^{2}} \gamma_{l m} \gamma_{ij} \tag3.12 \\
&+ \frac{1}{2} ~~ \frac{Y^{d}}{n} (h^{ab} \lambda_{ab})_{,d} \biggr\} \biggm|_{r=r_{+}}. \\ 
\endalign
$$
 One needs the full, readily computed expression for the Laplacian of $\Omega$ (given in the appendix) to evaluate the left hand side of Eq. (3.12) and, in particular, to extract the result for $\partial^3_r \Omega\bigm|_{r=r_{+}}$ but, when only angular derivatives of this Laplacian are needed, one can use instead the limiting form
$$
\left\{ \gamma^{ij} \tilde\nabla_i\tilde\nabla_j \Omega\right\}\biggm|_{r=r_{+}} \overset{\wedge}\to {=} \left\{ \frac{K}{2} ~ ~ h^{ab}\lambda_{ab}\right\}\biggm|_{r=r_{+}} . \tag3.13
$$

Attempting to go beyond this level and compute $\partial^4_r \Omega\bigm|_{r=r_{+}}$ however proves fruitless since, under further differentiation of the Hamiltonian constraint and evaluation at Scri, the coefficient of the fourth radial derivative of $\Omega$ at $\partial M$ is found to vanish identically.  On the other hand, given the results above for the first two radial derivatives of $\Omega$ at the boundary, it is straightforward to evaluate the traceless part of the Hessian of $\Omega$ at $\partial M$ and show that
$$\align
&\left\{ \Omega^{;rr} - \frac{1}{3} \gamma^{rr} (\gamma^{ij} \Omega_{;ij})\right\}\biggm|_{r=r_{+}} \overset{\wedge}\to {=} 0,\tag3.14 \\
&\left\{ \Omega^{;rd} - \frac{1}{3} ~ \gamma^{rd} (\gamma^{ij} \Omega_{;ij})\right\} \biggm|_{r=r_{+}} \overset{\wedge}\to {=} 0,
\endalign
$$
and
$$\align
&\bigl\{ \Omega^{;ef} - \frac{1}{3} \gamma^{ef} (\gamma^{ij}\Omega_{;ij}\bigr\}\bigm|_{r=r_{+}} \overset{\wedge}\to{=} \tag3.15 \\
& \frac{K}{3}\left[ (h^{ea} h^{fb} - \frac{1}{2} ~ h^{ef} h^{ab})\lambda_{ab}\right]\biggm|_{r=r_{+}}
\endalign
$$
where, as usual, we write $\Omega^{;ij}$ for $\gamma^{il}\gamma^{jm}\Omega_{;l m}$.
Note that the quantity in square brackets in the final equation is just the traceless part of the second fundamental form $\lambda_{ab}$ induced on $\partial M$.

By examining the angular components of Eq. (2.22) and comparing these with Eq. (3.15), we see that a necessary condition for regularity of the evolution equations at Scri will be the vanishing of the quantity
$$
\mu_\gamma \sigma^{ab} := \pi^{\tr\,{ab}} + \mu_\gamma (h^{ac}h^{bd} - \frac{1}{2} ~ h^{ab}h^{cd})\lambda_{cd} \tag3.16
$$
on this boundary.  This condition was derived by Andersson et al. in Ref. \cite{12} and identified geometrically there as equivalent to the well-known requirement that the shear of $\Cal I^+$ should vanish.  An examination of the remaining $(ri)$ components of Eq. (2.22), together with a comparison of these to the first and second of Eqs. (3.14), shows that we shall need the additional regularity conditions $\pi^{\tr\,{ri}} \overset{\wedge}\to {=} 0$ holding at Scri to avoid singularity there.  We shall see momentarily however that these apparently new restrictions are in fact forced by satisfaction of the momentum constraints and thus do not represent additional limitations upon the free data.

When the mean curvature is constant, as we have assumed, the momentum constraint, Eq. (2.7), takes the form 
$$
\Omega \left[ \pi^{\tr\,{ij}}_{\,\, ,j} + \tilde\Gamma^i_{l m} (\gamma) \pi^{\tr\,{l m}}\right] - 2 (\partial_j\Omega) \pi^{\tr\,{ij}} = 0. \tag3.17
$$
Evaluating this at $\partial M$, and recalling that $\Omega_{,r} \overset{\wedge}\to {=} - \frac{K}{3} ~ n$, leads immediately to the conclusion that
$$
\pi^{\tr\,{ri}} \overset{\wedge}\to {=} 0 \tag3.18
$$
which, as noted above, are necessary for regularity at the boundary.  Further important information results from computing the radial derivative of Eqs. (3.17) and evaluating the result at $\partial M$.  The `strong' form of the resulting equations (i.e., that attained before the boundary conditions $\pi^{\tr\,{ri}} \overset{\wedge}\to {=} 0$ have been enforced) is given by
$$\align
& \biggl\{ n^2(\frac{\pi^{\tr\,{ra}}}{n^{2}})_{,a} - \pi^{\tr\,{rr}}_{\,\, ,r} + \frac{\lambda_{ab}}{n} ~ \pi^{\tr\,{ab}} \tag3.19 \\
& - 2(\frac{n_{,r}}{n} + \frac{Y^{a}n_{\mid a}}{n} - \frac{n}{2} ~ h^{ab} \lambda_{ab}) \pi^{\tr\,{rr}} \\
& + 2 (\frac{n_{\mid a}}{n} + \frac{Y^{b}}{n} \lambda_{ab}) \pi^{\tr\,{ra}} \\
& + (\frac{n_{,r}}{n} + \frac{Y^{a}n_{\mid a}}{n} + \frac{1}{n} Y^a Y^b \lambda_{ab}) \pi^{\tr\,{rr}}\biggr\}\biggm|_{r=r_{+}} \overset{\wedge}\to {=} 0
\endalign
$$
and
$$\align
& \biggl\{ - ~ \pi^{\tr\,{cr}}_{\,\, ,r} + n^2 (\frac{\pi^{\tr\,{ca}}}{n^{2}})_{\mid a} - \frac{Y^{c}}{n} \lambda_{ab} \pi^{\tr\,{ab}} \tag3.20 \\
& - 2(\frac{n_{,r}}{n} + \frac{Y^{a}n_{\mid a}}{n} - \frac{n}{2} h^{ab} \lambda_{ab}) \pi^{\tr\,{cr}} \\
& + 2( - ~ nh^{cb}\lambda_{ab} - \frac{1}{n} Y^c Y^b\lambda_{ab} - \frac{Y^c}{n} n_{\mid a} + Y^c_{\mid a})\pi^{\tr\,{ra}} \\
& + ( - ~ \frac{Y^c}{n} n_{,r} + h^{cd} Y_{d,r} - \frac{Y^c}{n} Y^aY^b \lambda_{ab} \\
&\qquad - (h^{cd} + \frac{Y^cY^d}{n^2}) nn_{\mid d} - \frac{1}{2} (Y_aY^a)^{\mid c})\pi^{\tr\,{rr}}\biggr\}\biggm|_{r=r_{+}} \overset{\wedge}\to {=} 0.
\endalign
$$
When the boundary conditions are imposed however, these yield simply
$$
\{ \pi^{\tr\,{rr}}_{\,\, ,r}\}\bigm|_{r=r_{+}} \overset{\wedge}\to {=} \bigl\{\frac{\lambda_{ab}}{n} \pi^{\tr\,{ab}}\bigr\}\bigm|_{r=r_{+}} \tag3.21
$$
and
$$
\bigl\{ \pi^{\tr\,{rc}}_{\,\, ,r}\bigr\} \bigm|_{r=r_{+}} \overset{\wedge}\to {=} \bigl\{ n^2 \bigl(\frac{\pi^{\tr\,{ca}}}{n^2}\bigr)_{\mid a} - \frac{Y^c}{n} \lambda_{ab} \pi^{\tr\,{ab}}\bigr\}\bigm|_{r=r_{+}}. \tag3.22
$$
An application of L'Hospital's rule to the corresponding (apparently) singular terms $\frac{\pi^{\tr\,{ri}}}{\Omega}$ yields
$$\align
\lim\limits_{r\to r_+} \frac{\pi^{\tr\,{rr}}}{\Omega} &= \lim\limits_{r\to r_{+}} \bigl(\frac{\pi^{\tr\,{rr}}_{\,\, ,r}}{\Omega_{,r}}\bigr) \tag3.23 \\
&= - ~ \frac{3}{K} (\frac{\lambda_{ab}\pi^{\tr\,{ab}}}{n^2})\bigm|_{r=r_{+}}
\endalign
$$
and
$$\align
\lim\limits_{r\to r_+} \frac{\pi^{\tr\,{rc}}}{\Omega} &= \lim\limits_{r\to r_{+}} \bigl(\frac{\pi^{\tr\,{rc}}_{\,\, ,r}}{\Omega_{,r}}\bigr) = - ~ \frac{3}{K} \bigl[ n\bigl(\frac{\pi^{\tr\,{ca}}}{n^2}\bigr)_{\mid a} \tag3.24 \\
& -  \frac{Y^c}{n^2} ~ \lambda_{ab} \pi^{\tr\,{ab}}\bigr]\bigm|_{r=r_{+}} .
\endalign
$$

Using the foregoing results for the derivatives of $\Omega$ we find,
in the analogous way, regular boundary expressions for the remaining
(apparently) singular terms in Eq.~(2.22):
$$\align
&\bigl\{\frac{\Omega^{;rr} -\frac{1}{3}\gamma^{rr}(\gamma^{ij}\Omega_{;ij})}{\Omega}\bigr\}\bigm|_{r=r_{+}} \overset{\wedge}\to {=}  \\
&\bigl\{ - ~ \frac{h^{ab}}{n^3} n_{\vert ab} - \frac{2}{n^2} \bigl[\lambda_{ab}\lambda_{cd} h^{ac}h^{bd} + n^2 \tilde R^{rr}(\gamma) - \frac{1}{6} \tilde R(\gamma) \tag3.25 \\
&+ \frac{1}{6} \frac{\pi^{\tr\,{l j}}\pi^{\tr\,{im}}}{(\mu_\gamma)^2} \gamma_{l m}\gamma_{ij} -\frac{1}{4} (h^{ab}\lambda_{ab})^2 + \frac{1}{2n} Y^d (h^{ab} \lambda_{ab})_{,d} \\
&\qquad - \frac{1}{2n} (h^{ab} \lambda_{ab})_{,r}\bigr]\bigr\}\bigm|_{r=r_{+}},
\endalign
$$
$$\align
&\bigl\{ \frac{\Omega^{;rd}-\frac{1}{3}\gamma^{rd}(\gamma^{ij}\Omega_{;ij})}{\Omega}\bigr\}\bigm|_{r=r_{+}} \overset{\wedge}\to {=}  \\
&\bigl\{ \frac{n_{\vert c}}{n^2} h^{ad} h^{bc}\lambda_{ab} - \frac{h^{ad}}{n^2} (\frac{n_{\vert a}}{n})_{,r} + \frac{h^{ab}}{n^3} Y^d n_{\vert ab} \tag3.26 \\
&+ \frac{h^{ad}}{n^2} \bigl[\frac{n_{,r}}{n} - \frac{n}{2} h^{bc} \lambda_{bc}\bigr]_{,a} + \frac{2Y^d}{n^2} [\lambda_{ab}\lambda_{cf} h^{ac} h^{bf} + n^2 ~ \tilde R^{rr}(\gamma) \\
&- \frac{1}{6} \tilde R(\gamma) -\frac{1}{4} (h^{ab}\lambda_{ab})^2 + \frac{1}{6}\frac{\pi^{\tr\,{l j}}\pi^{\tr\,{im}}}{(\mu_\gamma)^{2}} \gamma_{l m}\gamma_{ij} + \frac{Y^f}{2n} (h^{ab}\lambda_{ab})_{,f} - \frac{1}{n} (\frac{h^{ab}\lambda_{ab}}{2})_{,r}\bigr]\bigr\}\bigm|_{r=r_{+}}
\endalign
$$
and
$$\align
&\bigl\{\frac{- ~ \frac{2}{3}\tilde NK\pi^{\tr\,{ef}}-2\mu_\gamma\tilde N(\tilde\nabla^e\tilde\nabla^f\Omega - 
\frac{1}{3}\gamma^{ef}(\gamma^{ij}\Omega_{;ij}))}{\Omega}\bigr\}\bigm|_{r=r_{+}} \overset{\wedge}\to{=}\\
&\bigl\{ \frac{2\tilde N}{n}\bigl[ \pi^{\tr\,{ef}}_{\,\, ,r} + n\mu_\gamma (\frac{2Y^eY^f}{n^2} -h^{ef})\bigl[\lambda_{ab}\lambda_{cd} h^{ac}h^{bd} + n^2 \tilde R^{rr}(\gamma) \\
&- \frac{1}{6} \tilde R(\gamma) - \frac{1}{2} (h^{ab}\lambda_{ab})^2 + \frac{1}{6} \frac{\pi^{\tr\,{l j}}\pi^{\tr\,{im}}}{(\mu_\gamma)^2} ~\gamma_{l m}\gamma_{ij} + \frac{Y^c}{2n} (h^{ab} \lambda_{ab})_{,c} \\
&- \frac{1}{2n} (h^{ab}\lambda_{ab})_{,r}\bigr] +\mu_\gamma \frac{n}{2} (h^{cd}\lambda_{cd})\bigl[-h^{ea}h^{fb}\lambda_{ab} + h^{ab} \lambda_{ab} \frac{Y^eY^f}{n^2}\tag3.27 \\
&- \frac{n_{\vert a}}{n^2} Y^e h^{fa} - \frac{n_{\vert a}}{n^2} Y^f h^{ea}\bigr] \\
& + \frac{\mu_\gamma n_{\vert c}}{n} [\lambda_{ab} h^{cb} h^{fa} Y^e + \lambda_{ab} h^{cb} Y^f h^{ea}]\\
&- \frac{n^2}{2} \mu_\gamma (h^{cd}\lambda_{cd})_{\vert a}\bigl(\frac{Y^eh^{fa}}{n^2} + \frac{Y^fh^{ea}}{n^2}\bigr) - \mu_\gamma (h^{ea}h^{fb} - h^{ab} \frac{Y^eY^f}{n^2}\bigr) n_{\vert ab} \\
&+ \frac{\partial}{\partial r} \bigl[ \mu_\gamma (h^{ea}h^{fb} -\frac{1}{2} h^{ef}h^{ab})\lambda_{ab}\bigr]\bigr]\bigr\}_{r=r_{+}}.
\endalign
$$
For the sake of generality, we have left Eqs. (3.25-3.27) in their `strong' form - they could be slightly further simplified by imposing $\pi^{\tr\,{ri}} \overset{\wedge}\to {=} 0$ at $\partial M$. One could also substitute the explicit expressions for $\tilde R^{ij}(\gamma)$ given in the appendix, into the above, but since there are other Ricci tensor contributions to the full $\pi^{\tr\,{ij}}$ evolution equations, it is more prudent to leave Eqs. (3.25-3.27) in their present form until the final evaluations are ready to be made.

Equations (2.9) and (2.13) can be combined in an obvious way to yield the following equation for the conformal lapse function
$$\align
& - \Omega (\gamma^{ij}\tilde N_{;ij}) - \tilde N(\gamma^{ij} \Omega_{;ij}) - \frac{\tilde R(\gamma)}{2} \Omega\tilde N \tag3.28 \\
& + 3\gamma^{ij}\tilde N_{,i} \Omega_{,j} + ~ \frac{3}{2} ~ \tilde N  ~\Omega ~ \frac{\pi^{\tr\,{im}} \pi^{\tr\,{jl}}}{(\mu_\gamma)^2} ~ \gamma_{ij} \gamma_{ml} \\
& = 0 .
\endalign
$$
Evaluating this at the boundary $\partial M$ gives
$$
\tilde N_{,r} \bigm|_{r=r_{+}} \overset{\wedge}\to {=} \bigl\{ Y^c \tilde N_{,c} - n\tilde N (\frac{h^{ab}\lambda_{ab}}{2})\bigr\}\bigm|_{r=r_{+}} \tag3.29
$$
where we have used Eq. (3.13) to reexpress the Laplacian of $\Omega$
at $r=r_{+}$.  Radially differentiating Eq. (3.28), we can derive an
expression for the Laplacian of $\tilde N$ at $\partial M$, from which a formula for $\tilde N_{,rr}\bigm|_{r=r_{+}}$ can be extracted.  Using these results it is straightforward to evaluate the $\mu_\gamma (\tilde\nabla^i\tilde\nabla^j\tilde N - \frac{1}{3} \gamma^{ij} \gamma^{mn} \tilde\nabla_m\tilde\nabla_n\tilde N)$ contributions to the field equations at $\partial M$. One cannot compute higher radial derivatives of $\tilde N$ in this way however since that would necessitate knowledge of the values of higher than third radial derivatives of $\Omega$ at the boundary. In every case however, the Taylor expansion techniques allow the computation of precisely what is needed for evaluation of the evolution equations at Scri. We shall present these equations explicitly in the next section and show how they imply preservation of the regularity conditions $\pi^{\tr\,{ri}} \overset{\wedge}\to{=} 0$ and $\pi^{\tr\,{ab}} + \mu_\gamma (h^{ac}h^{bd} - \frac{1}{2} h^{ab} h^{cd})\lambda_{cd} \overset{\wedge}\to {=} 0$ at the conformal boundary.

Though we have assumed constant mean curvature throughout, it would not be difficult to incorporate the contributions of variable mean curvature. For example, letting $\tau = \frac{\tr_g\pi}{\mu_g} = \frac{g_{ij}\pi^{ij}}{\mu_g}$, and allowing $\tau$ to be variable, one gets for the momentum constraint
$$
\Omega \tilde\nabla_j \pi^{\tr\,{mj}} - 2\Omega_{,j} \pi^{\tr\,{mj}} + \frac{1}{3} \mu_\gamma \gamma^{im} \partial_i \tau = 0. \tag3.30 
$$
This leads to the modified regularity constraint
$$
\bigl\{ \pi^{\tr\,{ri}} - \bigl( \frac{\mu_\gamma}{n\tau}\bigr) \gamma^{ij} \partial_{j} \tau\bigr\}\bigm|_{r=r_{+}} \overset{\wedge}\to {=} 0. \tag3.31
$$
Similar modifications are implied for $\pi^{\tr\,{ri}}_{\,\, ,r}$ and for the radial derivatives of $\Omega$ and $\tilde N$ at Scri. We do not, however, anticipate that such modifications would interfere significantly with the main conclusions derived herein. Instead, we believe that the CMC condition plays a rather inessential role in our analysis but we prefer to retain it because of the associated, partial decoupling of the constraints that it allows.  

We also note that, whereas our full constrained evolution proposal (as sketched in section II) entailed a constant scalar curvature normalization and harmonic coordinate conditions for the metric $\gamma_{ij}$, the corresponding (non-degenerate) elliptic equations for the shift vector $X^i \frac{\partial}{\partial x^i}$ (Eq. (2.15)) and the function $\Gamma$ (Eq. (2.12)) played no role in the above analysis. Thus we do not believe that these particular choices are at all essential for the central conclusions derived herein. Rather we expect that many alternative formulations of the field equations could be adapted to putting the outer boundary at Scri and that our specific proposal is just one of many feasible possibilities for doing so. On the other hand, we also believe that many of our calculations have a rather `universal' character and will be applicable to a variety of alternative formulations which adopt different gauge conditions or normalizations from the ones we have chosen.

We remark here that one of the main advantages of preserving, as we have, the strong forms of the relevant equations is that these lend themselves to further generalization (through modification of the gauge or normalization conditions or the introduction of material sources) by the straightforward computation of the additional terms necessitated by the desired modification.  By contrast the information lost in passing to the weak forms of these equations would almost surely necessitate a rederivation of most of the relevant formulas from scratch.

\head IV. Preservation of Regularity Conditions \endhead
\vskip .10in 

In this section, we fit together the various contributions to the $\pi^{\tr\,{ij}}$ evolution equations derived above and show that they imply
$$
\frac{\partial\pi^{\tr\,{ri}}}{\partial t} \overset{\wedge}\to {=} 0, \,\,  \frac{\partial}{\partial t} (\pi^{\tr\,{ab}} + \mu_\gamma (h^{ac} h^{bd} - \frac{1}{2} ~ h^{ab} h^{cd})\lambda_{cd}) \overset{\wedge}\to {=} 0 \tag4.1
$$
independently of any further restriction upon the geometrical data at Scri (i.e., without the need for any additional constraint on the boundary values of\newline $\{ h_{ab}, n, Y^a, \tilde N, Z^a, \lambda_{ab}, \Gamma ~ \text{or} ~ \tilde R_{ij} (\gamma)\}$).  We do this by first simply evaluating the `strong' form of the relevant evolution equations and then noting, by inspection, that when the regularity constraints, $\{ \pi^{\tr\,{ri}} \overset{\wedge}\to {=} 0, \,\, \mu_\gamma \sigma^{ab} := \pi^{\tr\,{ab}} + \mu_\gamma (h^{ac}h^{bd} - \frac{1}{2} h^{ab} h^{cd})\lambda_{cd} \overset{\wedge}\to {=} 0\}$, are enforced these evolution equations reduce to the (trivial) forms given above.

There is however, a subtlety in this seemingly straightforward procedure that we wish to address at the outset. This concerns the logical significance of the `strong' forms of some of the relevant equations and the justification for replacing them with their corresponding `weakened' forms in deriving the main results. To see the issue at hand, consider for a moment our derivation of the expressions (3.23 and 3.24) for the boundary values of the apparently singular terms 
$\frac{\pi^{\tr\,{ri}}}{\Omega}$ using L'Hospital's rule. We chose, for simplicity, to present those results in their `weak' forms by dropping all those contributions from equations (3.19 and 3.20) that vanish at Scri by virtue of the regularity conditions even though the retention of such terms would still have yielded a completely regular result.  The strong forms of such evaluations, though perfectly regular at Scri, entail a certain logical contradiction -- one assumes the regularity conditions hold in order to extract finite limits from otherwise singular expressions but then, after applying L'Hospital's rule, leaves in the $\pi^{\tr\,{ri}}$ and $\sigma^{ab}$ contributions as though the regularity constraints were actually being relaxed.  For other purely regular terms in the evolution equations, no such contradictory assumptions are involved in their evaluation and so their strong forms seem less problematic.

Of course, one is always free to drop all weakly vanishing terms and restore logical consistency so no harm is done in retaining such terms in early stages of the calculation.  We suspect however that in some eventual, deeper mathematical study of these issues, it may be important to know the actual structure of such weakly vanishing `forcing' terms and so we have uniformly retained them in the derivations reported below.

The reader may well wonder, however, whether we are not then obligated to prove a strong version of conservation of the regularity conditions by establishing a form of hyperbolicity of the boundary evolution equations or (essentially equivalently) deriving suitable `energy' estimates to show that these equations have only the trivial solution for vanishing initial data.  Here, however, the illogic in the derivation of such strong equations comes to the foreground.  If one imagines that the regularity conditions are not necessarily enforced at Scri (as would be implicit in the use of the strong form of the evolution equations) then one has no logical right to make contradictory assumptions for the 
`evaluation' (using L'Hospital's rule) of genuinely singular limits such as $\lim\limits_{r\nearrow r_{+}}
 \frac{\pi^{\tr\,{ri}}}{\Omega}$ and $\lim\limits_{r\nearrow r_{+}} \frac{\sigma^{ab}}{\Omega}$.  On the other hand, the explicit contributions of these seemingly problematic expressions to the corresponding evolution equations, namely the terms
$$
- \frac{2}{3} \frac{\tilde N}{\Omega} K \pi^{\tr\,{ri}} ~ \text{and} ~  -\frac{2}{3} \frac{\tilde N}{\Omega} K \mu_\gamma \sigma^{ab},
$$
take the forms of `frictional forcing' terms that (having good sign for evolution towards the future) seek to drive the quantities $\pi^{\tr\,{ri}}$ and $\sigma^{ab}$ towards vanishing values with effective frictional coefficients that diverge like $\Omega^{-1}$ as Scri is approached. Intuitively speaking then, the regularity variables $\{ \pi^{\tr\,{ri}}, \sigma^{ab}\}$ feel precisely zero force when they lie on the `constraint submanifold' defined by the regularity conditions $\pi^{\tr\,{ri}} \overset{\wedge}\to {=} 0, ~ \sigma^{ab} \overset{\wedge}\to {=} 0$ but would be subject to an infinitely strong `frictional' restoring force if they tried to `float off' this manifold. Finding a precise mathematical characterization of this state of affairs might prove to be a subtle task, but we are encouraged that the indicators that we have identified all point towards stable preservation of the regularity conditions at Scri.

In the previous section, we showed how to evaluate the most problematic contributions to the evolution equations at Scri. The remaining terms (i.e., those involving the Lie derivative of $\pi^{\tr\,{ij}}$, non-singular algebraic expressions and terms involving the Ricci tensor  of $\gamma_{ij}$) are straightforward evaluations that require little comment. We recall however, from section II, that the angular components of the shift vector, $X^i \frac{\partial}{\partial x^i}$, have the following form at Scri
$$
X^d\bigm|_{r=r_{+}} \overset{\wedge}\to {=} \bigl\{\frac{3}{K} \tilde N \gamma^{dj} \Omega_{,j} + Z^d\bigr\}
\bigm|_{r=r_{+}} \tag4.2
$$
or equivalently, using foregoing results on the behavior of $\Omega$ at $\partial M$,
$$
X^d \bigm|_{r=r_{+}} \overset{\wedge}\to {=}  \bigl\{ \frac{\tilde N}{n} Y^d + Z^d\bigr\}\bigm|_{r=r_{+}}. \tag4.3
$$
Accordingly, Lie derivatives with respect to the $2$-dimensional vector fields\newline
$^{(2)}\tilde X := X^a \frac{\partial}{\partial x^a}, ~ ^{(2)}\tilde Z := Z^a \frac{\partial}{\partial x^a}$ and $^{(2)}\tilde Y := Y^a \frac{\partial}{\partial x^a}$ are expressed as $\Cal L_{{}^{(2)} \tilde X}, \Cal L_{{}^{(2)} \tilde Z}$ and 
$\Cal L_{{}^{(2)} \tilde Y}$ respectively.

Assembling the various elements of the (strong forms of the) $\pi^{\tr\,{ri}}$ evolution equations, one thence obtains
$$\align
&\bigl\{ \frac{\partial}{\partial t} \pi^{\tr\,{rr}}\bigr\}\bigm|_{r=r_{+}} \overset{\wedge}\to {=} \bigl\{ -X^{r}_{\,\, ,r} \pi^{\tr\,{rr}} - 2X^r_{,a} \pi^{\tr\,{ra}} + (X^a\pi^{\tr\,{rr}})_{,a} \tag4.4 \\
&- \frac{2\tilde N}{\mu_\gamma} \bigl[(\pi^{\tr\,{rr}})^2 (n^2 +Y_cY^c)+ 2 Y_a \pi^{\tr\,{ra}}\pi^{\tr\,{rr}} + h_{ab} \pi^{\tr\,{ra}}\pi^{\tr\,{rb}}\bigr] \\
&+ \frac{\tilde N}{n}\bigl[ \pi^{\tr\,{ra}}_{\,\, ,a} + \frac{2Y^b}{n} \lambda_{ab} \pi^{\tr\,{ra}} + \bigl( -\frac{n_{,r}}{n} - \frac{Y^cn_{\vert c}}{n} + \frac{1}{n} Y^aY^b\lambda_{ab} + nh^{ab} \lambda_{ab}\bigr)\pi^{\tr\,{rr}}\bigr] \\
&+ \frac{\tilde N}{n^2} \lambda_{ab} \bigl[\pi^{\tr\,{ab}} + \mu_\gamma (h^{ac}h^{bd} - \frac{1}{2} h^{ab} h^{cd})\lambda_{cd}\bigr]\bigr\}\bigm|_{r=r_{+}},
\endalign
$$
$$\align
& \bigl\{ \frac{\partial}{\partial t} \pi^{\tr\,{rd}}\bigr\}\bigm|_{r=r_{+}} \overset{\wedge}\to {=} \bigl\{ - ~~
 \frac{\tilde N Y^d}{n^2} \lambda_{ab} \bigl[ \pi^{\tr\,{ab}} + \mu_\gamma (h^{ac}h^{bf} -\frac{1}{2} h^{ab}h^{cf})\lambda_{cf}\bigr]\tag4.5 \\
&- X^d_{,r} \pi^{\tr\,{rr}} - X^d_{,a} \pi^{\tr\,{ra}} + (X^a \pi^{\tr\,{rd}})_{,a} \\
& -\frac{2\tilde N}{\mu_\gamma} \bigl[ \pi^{\tr\,{rr}} \pi^{\tr\,{rd}} (n^2 + Y_cY^c) + \pi^{\tr\,{rr}} \pi^{\tr\,{da}}Y_a \\
&\qquad + \pi^{\tr\,{ra}} \pi^{\tr\,{dr}} Y_a + h_{ab} \pi^{\tr\,{ra}} \pi^{\tr\,{bd}}\bigr] \\
& + \frac{\tilde N}{n} \bigl[ - ~ ~ 2 (\frac{n_{,r}}{n} + \frac{Y^an_{\vert a}}{n} - \frac{n}{2} h^{ab} \lambda_{ab}\bigr) \pi^{\tr\,{dr}} \\
& + 2\pi^{\tr\,{ra}}(- nh^{db} \lambda_{ab} - \frac{1}{n} Y^d Y^b \lambda_{ab} - \frac{Y^d}{n} n_{\vert a} + Y^d_{\vert a}) \\
& + \pi^{\tr\,{rr}}\bigl(- ~ ~\frac{Y^d}{n} n_{,r} + h^{df} Y_{f,r} - \frac{Y^d}{n} Y^a Y^b \lambda_{ab} \\
&\qquad - (h^{df} + \frac{Y^dY^f}{n^2})nn_{\vert f} - \frac{1}{2} (Y_aY^a)^{\vert d}\bigr)\bigr] \\
&+ \tilde N n^2\bigl[\frac{\pi^{\tr\,{dc}} + \mu_\gamma(h^{ad}h^{bc}-\frac{1}{2} h^{cd}h^{ab})\lambda_{ab}}{n^3}\bigr]_{\vert c}  \\
& + \frac{\tilde N_{\vert c}}{n} \bigl[ \pi^{\tr\,{cd}} + \mu_\gamma (h^{ad}h^{bc} -\frac{1}{2} h^{cd} h^{ab})\lambda_{ab}\bigr]\bigr\}\bigm|_{r=r_{+}} .
\endalign
$$
Proceeding in the same way, one finds for the evolution equation of the densitized shear,
$\pi^{\tr\,{ef}} + \mu_\gamma (h^{ea} h^{fb} - \frac{1}{2} h^{ef} h^{ab}) \lambda_{ab}$,

\newpage

$$
\frac{\partial}{\partial t}\bigl\{ (\pi^{\tr\,{ef}} + \mu_\gamma (h^{ea}h^{fb} -\frac{1}{2} h^{ef} h^{ab})\lambda_{ab})\bigr\}\bigm|_{r=r_{+}} \tag4.6 
$$
$$\align
&\overset{\wedge}\to{=} \{\frac {-2\tilde N}{\mu_\gamma}[\pi^{\tr\,{er}}\pi^{\tr\,{fr}} (n^2+ Y_cY^c) + \pi^{\tr\,{er}}\pi^{\tr\,{fa}} Y_a + \pi^{\tr\,{ea}} \pi^{\tr\,{fr}} Y_a + \pi^{\tr\,{ea}} \pi^{\tr\,{fb}} h_{ab}] \\
&- \mu_\gamma \tilde N(Y^eh^{af} + Y^f h^{ae})\bigl[\bigl(\frac{\lambda_{ac}}{n}\bigr)^{\mid c} - \frac{1}{2} \bigr( \frac{h^{bd}\lambda_{bd}}{n}\bigl)_{\mid a}\bigr] \\
&+ \tilde N \mu_\gamma \bigl(\frac{Y^eY^f}{n^2} - \frac{h^{ef}}{2}\bigr) \bigl[ - \frac{1}{2} (h^{ab}\lambda_{ab})^2 + h^{ab}h^{cd} \lambda_{ad}\lambda_{bc}\bigr] \\ 
&+ \Cal L_{{}^{(2)} \tilde X} \bigl[ \pi^{\tr\,{ef}} + \mu_\gamma (h^{ea} h^{fb} - \frac{1}{2} h^{ef}h^{ab})\lambda_{ab}\bigr] \\
&+ X^r_{\, , r}(\pi^{\tr\,{ef}} + \mu_\gamma (h^{ea}h^{fb} - \frac{1}{2} h^{ef}h^{ab})\lambda_{ab}) - X^e_{\, ,r} \pi^{\tr\,{rf}} - X^f_{\, ,r} \pi^{\tr\,{er}} \\
& +\mu_\gamma (h^{ea} h^{fb} - \frac{1}{2} h^{ef} h^{ab})\bigl[ - 2\tilde N h^{cd} \lambda_{ad}\lambda_{bc} + \tilde N \lambda_{ab} \bigl(\frac{h^{cd}\lambda_{cd}}{2}\bigr) \\
&- \lambda_{ab}\frac{\tilde N}{\mu_\gamma} n^2 \pi^{\tr\,{rr}} + \frac{1}{2n} \bigl[ 2Y_b \frac{\tilde N}{\mu_\gamma} n^2 \pi^{\tr\,{rr}} + \frac{2\tilde N}{\mu_\gamma} n^2 h_{cb}\pi^{\tr\,{rc}}\bigr]_{\mid a} \\
&+ \frac{1}{2n} \bigl[ 2Y_a \frac{\tilde N}{\mu_\gamma} n^2 \pi^{\tr\,{rr}} + \frac{2\tilde N}{\mu_\gamma} n^2 h_{ac} \pi^{\tr\,{rc}}\bigr]_{\mid b} \\
& - \frac{1}{2n} \frac{\partial}{\partial r} \bigl\{ \frac{2\tilde N}{\mu_\gamma} \bigl[ Y_aY_b \pi^{\tr\,{rr}} + Y_a h_{bd} \pi^{\tr\,{rd}} + Y_b h_{ad} \pi^{\tr\,{rd}}\bigr]\bigr\} \\
&+ \frac{1}{2n} \Cal L_{{}^{(2)} \tilde Y} \bigl\{ \frac{2\tilde N}{\mu_\gamma} (Y_aY_b \pi^{\tr\,{rr}} + Y_a h_{bd}\pi^{\tr\,{rd}} + Y_bh_{ad} \pi^{\tr\,{rd}})\bigr\}\bigr] \\
&+ \mu_\gamma \lambda_{ab} (-h^{ec} h^{ad} h^{fb} - h^{ea}h^{fc} h^{bd} + \frac{1}{2} h^{ec} h^{fd} h^{ab} + \frac{1}{2} h^{ef} h^{ac} h^{bd})\\
&\qquad \times \frac{2\tilde N}{\mu_\gamma} (Y_c Y_d \pi^{\tr\,{rr}} + Y_c h_{dm} \pi^{\tr\,{rm}} + Y_d h_{cm} \pi^{\tr\,{rm}}) \\
&+ \frac{\mu_\gamma}{n} ~ \frac{\partial}{\partial r} \bigl[ \frac{1}{2} \frac{\tilde N}{\mu_\gamma} h^{ef}\bigr] h_{cd} \pi^{\tr\,{cd}} + \frac{\tilde N}{n} \frac{1}{2} h^{ef} \frac{\partial}{\partial r} [ h_{cd} \pi^{\tr\,{cd}}] \\
&+ \pi^{\tr\,{ef}}(\frac{\tilde N}{2} h^{ab} \lambda_{ab}) + \frac{\mu_\gamma}{n} \Cal L_{{(2)}_{\tilde Y}} \bigl[ - ~ ~\frac{1}{2} \frac{\tilde N}{\mu_\gamma} h^{ef} h_{cd} \pi^{\tr\,{cd}}\bigr] \\
&+ 2\tilde N \bigl[ \frac{1}{2} \pi^{\tr\,{ef}} (h^{ab} \lambda_{ab}) - \frac{1}{2} \lambda^{ef} h_{ab} \pi^{\tr\,{ab}} + \mu_\gamma (2\lambda^{ea} \lambda^f_a - \frac{1}{2} \lambda^{ef} h^{ab} \lambda_{ab} - \frac{1}{2} h^{ef} \lambda^{ab} \lambda_{ab})\bigr] \\
&+ \frac{\tilde N}{n} \frac{\mu_{\gamma ,r}}{\mu_\gamma} \bigl(
\pi^{\tr\,{ef}} + \mu_\gamma (h^{ea} h^{fb} - \frac{1}{2} h^{ef}
h^{ab})\lambda_{ab}\bigr) \\& + \frac{\mu_{\gamma}\tilde N}{n} \Cal L_{{(2)}_{\tilde Y}}
\bigl[ \frac{\pi^{\tr\,{ef}}}{\mu_\gamma} + (h^{ea} h^{fb} - \frac{1}{2} h^{ef} h^{ab})\lambda_{ab}\bigr]\bigr\}\bigm|_{r=r_+}\\
\endalign 
$$

For the first two equations (4.4) and (4.5) it is clear by inspection that the right hand sides vanish `weakly' (i.e., vanish when the regularity constraints $\pi^{\tr\,{ri}} \overset{\wedge}\to {=} 0$ and $\mu_\gamma \sigma^{ab} := \bigl[ \pi^{\tr\,{ab}} + \mu_\gamma (h^{ac}h^{bd} - \frac{1}{2} h^{ab}h^{cd})\lambda_{cd}\bigr] \overset{\wedge}\to {=} 0
$ are enforced at $\partial M$). To verify that the right hand side of the third equation (4.6) also vanishes weakly, requires a little work.  For this purpose it is important to note that since the $3$-dimensional trace, $\gamma_{ij} \pi^{\tr\,{ij}}$, of $\pi^{\tr\,{ij}}$ vanishes identically, one has
$$\align
\gamma_{ij} \pi^{\tr\,{ij}} &= h_{ab} \pi^{\tr\,{ab}} + (n^2 + Y_cY^c)\pi^{\tr\,{rr}} + 2 Y_a \pi^{\tr\,{ra}} \tag4.7 \\
& = 0
\endalign
$$
and thus finds that the $2$-dimensional trace, $h_{ab} \pi^{\tr\,{ab}}$, of $\pi^{\tr\,{ab}}$ vanishes weakly at $\partial M$.  It is also useful to recall that an arbitrary $2$-dimensional, traceless, symmetric tensor $s^{ab}$ satisfies the identity
$$
s^{ac}s^{bd} h_{cd} = \frac{1}{2} h^{ab} (h_{ce}h_{df} s^{cd}s^{ef}) \tag4.8
$$
when $s^{ab}= s^{ba}$ and $h_{ab}s^{ab} = 0$.

Thus equations (4.4 - 4.6) reduce weakly to equations (4.1) and so
imply the preservation of the regularity conditions throughout the
evolution at Scri. It is important to note, especially if alternative
gauge conditions, normalizations or formulations of the field
equations are under consideration, that our results involve no
implicit restriction upon the boundary values of the geometric data
$\{ h_{ab}, n, Y^a, \tilde N, Z^a, \lambda_{ab}, \Gamma ~ \text{or} ~
\tilde R_{ij}(\gamma)\}$.  Our main tools have been simply the
enforcement of the Hamiltonian and momentum constraints together with
the straightforward application of Taylor expansions and L'Hospital's
rule for the evaluation of apparently singular terms at $\partial M$.
Although we have also imposed the CMC slicing condition (and peripherally, the spatial harmonic gauge condition), we do not believe that these were at all essential for our principal conclusions and that one could relax them without harmful effect.
\vskip .15in

\head V. Alternative Evaluation of Singular Terms\endhead
\vskip .10in

As is well-known \cite{14, 15} the electric components of the Weyl tensor, for a solution of the vacuum field equations, can be expressed in terms of the physical Cauchy data $(g_{ij}, K_{ij})$ as
$$
E_{ij} = R_{ij} (g) - K_{im} K^m_{\,\,\, j} + K_{ij} \tr_g K \tag5.1
$$
where $K^m_{\,\,\, j} = g^{ml} K_{l j}$ and $\tr_gK = g^{kl} K_{kl}$.  Looking at this equation though, one may well wonder why, since the Weyl tensor is conformally invariant, the above formula for $E_{ij}$ is not.  The answer in part is that the above expression does not correspond to the fundamental formula for $E_{ij}$ but rather to a representation of it that has been transformed, through the application of the ADM field equations, to a form in which time derivatives have been eliminated in favor of spatial ones. Thus the above expression for $E_{ij}$ has, in effect, inherited the failure of the ADM equations to be conformally invariant.  But this same lack of conformal invariance in the ADM equations is precisely the feature which led, upon conformal rescaling, to the appearance of the singular terms that we have been concerned with. Thus it should perhaps not be surprising to find that there is a close relationship between those singular terms and the conformal transformation properties of the above expression for $E_{ij}$.

To see this explicitly, let us define
$$
\Cal E_m^{\,\,\, j} := \mu_g E_m^{\,\,\, j} = \mu_g g^{jl} E_{ml} \tag5.2
$$
and reexpress the physical variables in terms of the conformal ones introduced in section II.  Even though the trace of $\Cal E_m^{\,\,\, j}$ vanishes by virtue of the Hamiltonian constraint, it is convenient to write the resulting formula in terms of the explicitly trace-free quantity $\Cal E_m^{\,\,\, j} - \frac{1}{3} \delta^j_m \Cal E_k^{\, k}$ since this facilitates comparison with our earlier derivations.  Using the well-known conformal transformation properties of the Ricci tensor, and introducing the gravitational momentum variables $\pi^{\tr\,{ij}}$ in favor of $K^{tr}_{ij}$, one easily arrives at:
$$\align
& -\frac{2}{3} \frac{K\tilde N}{\Omega} \pi^{\tr\,{ij}} - 2\tilde N \mu_\gamma \bigl(\frac{\tilde\nabla^i\tilde\nabla^j\Omega}{\Omega} - \frac{1}{3} \frac{\gamma^{ij} \tilde\nabla_k\tilde\nabla^k\Omega}{\Omega}\bigr) \tag5.3 \\
&= 2\tilde N \mu_\gamma (\tilde R^{ij} (\gamma) - \frac{1}{3} \gamma^{ij} \tilde R(\gamma)) \\
& -\frac{2\tilde N}{\mu_\gamma} \bigl[ \gamma_{ml} \pi^{\tr\,{il}} \pi^{\tr\,{jm}} - \frac{1}{3} \gamma^{ij} \gamma_{ml}\gamma_{nk} \pi^{\tr\,{kl}} \pi^{\tr\,{mn}}\bigr] \\
& - 2\tilde N \Omega \gamma^{im} (\Cal E_m^{~ ~ j} - \frac{1}{3} \delta^j_m \Cal E_k^{~ ~ k} ).
\endalign
$$
The left hand side of this equation consists of precisely the apparently singular terms in the $\pi^{\tr\,{ij}}$ evolution equation (2.22) whereas the right hand side consists of purely regular terms provided that $\Omega\Cal E_m^{~ ~ ~j}$ is regular at $\Cal I^+$.

The reader may wonder however, whether we have hidden some singular behavior in the notation by choosing the mixed, densitized form $\Cal E_m^{~ ~ ~ j}$ to represent the electric components of the Weyl tensor. That we have not done so however, may be seen by writing down the evolution and constraint equations satisfied by $\Cal E_m^{~ ~ ~ j}$ and the corresponding magnetic components $\Cal B_m^{~ ~ ~ j}$, expressible in terms of ADM variables as
$$\align
\Cal B_m^{~ ~ ~ j} &:=  ~ \mu_g B_m^{ ~ ~ ~ ~ j} = ~ \mu_g g^{jl} B_{ml} \tag5.4 \\
&= ~ \frac{1}{2} ~ ~ \varepsilon^{il j} (\nabla_i (g) K_{l m} - \nabla_l (g) K_{im})
\endalign
$$
or, upon assuming that $\tr_g K =$ constant as above and reexpressing $K_{ij}$ in terms of the gravitational momentum, as
$$
\Cal B_m^{ ~ ~ ~ j} = \frac{\varepsilon^{il j}}{\mu_g} ~ g_{mn} ~ ~ g_{is} \bigl( \nabla_l (g) ~ \pi^{\tr\,{ns}}\bigr).
\tag5.5
$$ 
These Maxwell-like evolution and constraint equations, derivable directly from the defining formulas for $\Cal E_m^{~ ~ ~ j}$ and $\Cal B_m^{~ ~ ~ j}$ through an application of the ADM equations, can be expressed as:  
$$\align
& \frac{\partial}{\partial t} ~ \Cal E_j^{~ ~ ~ l} = (\Cal L_{\tilde X} \Cal E )_j^{~ ~ ~ l} \tag5.6 \\
&+ \mu_\gamma \gamma^{il} \{ - \varepsilon_{rsj} \tilde N_{,m} \gamma^{rm} \Cal B_i^{~ ~ ~ s} \\
&\qquad - \varepsilon_{rsi} ~ \tilde N_{,m} \gamma^{rm} \Cal B_j^{~ ~ ~ s} \\
&\qquad+ \frac{\tilde N}{2} [- \varepsilon_{rsj} \gamma^{rm} \tilde\nabla_m (\gamma) \Cal B_i^{~ ~ ~ s} - \varepsilon_{rsi} \gamma^{rm} ~ \tilde\nabla_m (\gamma) \Cal B_j^{~ ~ ~ s}] \} \\
&+ \frac{5}{2} ~ \frac{\tilde N\gamma_{jk}}{\mu_\gamma} ~ \Cal E_m^{~ ~ ~ l} \pi^{\tr\,{km}} \\
&- ~ ~ \frac{\tilde N\gamma_{km}}{\mu_\gamma} ~ \delta^l_j \pi^{\tr\,{kn}} \Cal E_n^{~ ~ ~ m} \\
&+ \frac{1}{2} ~ \frac{\tilde N}{\mu_\gamma} ~ \gamma_{km} \pi^{\tr\,{kl}} ~ \Cal E_j^{~ ~ ~ m} , \\ 
\\
& \frac{\partial}{\partial t} ~ \Cal B_j^{~ ~ ~ l} = (\Cal L_{\tilde X} \Cal B)_j^{~ ~ ~ l} \\
&+ \mu_\gamma \gamma^{il} \{ \varepsilon_{rsj} ~ \tilde N_{,m} \gamma^{rm} \Cal E_i^{~ ~ ~ s} + ~ \varepsilon_{rsi} ~ \tilde N_{,m} \gamma^{rm} \Cal E_j^{~ ~ ~ s} \\
&\qquad+ \frac{\tilde N}{2} [\varepsilon_{rsj} \gamma^{rm} \tilde\nabla_m(\gamma) \Cal E_i^{~ ~ ~ s} + ~ \varepsilon_{rsi} \gamma^{rm} \tilde\nabla_m(\gamma) \Cal E_j^{~ ~ ~ s}] \} \\
&+ ~ \frac{5}{2} ~ \frac{\tilde N \gamma_{jk}}{\mu_\gamma} ~ \Cal B_m^{~ ~ ~ l} \pi^{\tr\,{km}} \\
& - ~ \frac{\tilde N \gamma_{km}}{\mu_\gamma} ~ \delta_j^{~ ~ ~ l} \pi^{\tr\,{kn}} ~ \Cal B_n^{~ ~ ~ m} \\
& + ~ \frac{1}{2} ~ \frac{\tilde N \gamma_{km}} {\mu_\gamma} ~ \pi^{\tr\,{kl}} ~ \Cal B_j^{~ ~ ~ m} , \tag5.7 \\
\\
& \tilde\nabla_j(\gamma) \Cal E_m^{~ ~ ~ j} = - ~ \varepsilon_{ml r} \pi^{\tr\,{l s}} ~ \Cal B_s^{~ ~ ~ r}, \tag5.8 \\
\\
& \tilde\nabla_j(\gamma) ~ \Cal B_m^{~ ~ ~ j} =  + ~
\varepsilon_{ml r} \pi^{\tr\,{l s}} ~ \Cal E_s^{~ ~ ~ r}, \tag5.9 
\endalign
$$
where $\tilde X := X^i \frac{\partial}{\partial x^i}$ and
$\Cal{L}_{\tilde X}$ signifies Lie differentiation with respect
to $\tilde X$.

Though we have written the above in terms of the conformal metric
variables, we could just as well have replaced them by the physical
metric components since it is easy to see by inspection that all of
Eqns. (5.6-5.9) are conformally invariant. For this reason, therefore, the possibility that some regular null hypersurface in the unphysical geometry may in fact play the role of future null infinity in the physical geometry is completely invisible to the quantities 
$(\Cal E_m^{~ ~ ~ j}, \Cal B_m^{~ ~ ~ j})$, which evolve without any
interaction with the conformal factor $\Omega$. Thus, unless some
singular behavior is put `by hand' into the initial conditions for
$\Cal E_m^{~ ~ ~ j}$ and $\Cal B_m^{~ ~ ~ j}$ at $\Cal I^+$, there is
no reason to anticipate that these quantities will blow up along null
infinity, at least for sufficiently short time intervals during the
evolution. This plausibility argument that singular boundary behavior
for $(\Cal E_m^{~ ~ ~ j}, \Cal B_m^{~ ~ ~ j})$ should not be put in
'by hand' can, however, be strengthened to a mathematical proof (due originally to Penrose \cite{2}) that sufficiently smooth conformal compactifications must necessarily have vanishing unphysical Weyl tensors at Scri.  As we shall see below, this result in turn implies that $\Omega\Cal E_m^{~ ~ ~ l}$ and $\Omega\Cal B_m^{~ ~ ~ l}$ must both vanish at Scri.

Thus the issue of whether $\Omega\Cal E_m^{~ ~ ~ l}$ should be assumed to vanish at Scri hinges ultimately upon the `reasonableness' of Penrose's smoothness hypotheses (which require $C^3$-differentiability of the conformal metric out to the boundary). But this question opens the Pandora's box of concerns as to whether one should allow so-called `polyhomogeneous' (or polylogarithmic) boundary behavior as a `natural' alternative to Penrose's smoothness requirements.  The issue of whether polyhomogeneous Scris are actually needed for sufficient physical generality is still an open question upon which we currently do not wish to express a definite opinion.  For simplicity therefore, we have adopted the traditional Penrose viewpoint that null infinity should be sufficiently smooth that the corresponding argument for the vanishing of the unphysical Weyl tensor at Scri can be applied.  As we shall discuss more fully below however, most of our calculations are actually compatible with the possibility that Scri could be polyhomogeneous instead of smooth in the Penrose sense.

Assuming therefore that the quantities $\Omega\Cal E_m^{~ ~ ~ j}$ vanish at $\Cal I^+$ we obtain from Eq. (5.3) an independent expression for the apparently singular terms in the evolution equations for $\pi^{\tr\,{ij}}$ in terms of regular quantities. How do these compare with our previous calculations? It is straightforward to verify that, for the $ri$ components of these expressions, the two formulas agree weakly (i.e., upon satisfaction of the regularity conditions) at the conformal boundary but not strongly. A direct comparison of the expressions for the angular components however, is not possible since our previous calculation, through its application of L'Hospital's rule, required the boundary values of $\pi^{\tr\,{ab}}_{~ ~ ,r}$ whereas the Weyl formula given above involves no derivatives  of $\pi^{\tr\,{ab}}$ at $\Cal I^+$.
L'Hospital's rule was of course also used for the evaluation of the $ri$ components but there we could use the momentum constraints to compute the corresponding radial derivatives, $\pi^{\tr\,{ri}}_{~ ~ ~ ,r}$ at $\Cal I^+$.
We did not need boundary expressions for the quantities $\pi^{\tr\,{ab}}_{~ ~ ~ ,r}$ in the calculation to show that the  vanishing of the shear of Scri is preserved since the contributions of these radial derivatives (including the Lie derivative of $\pi^{\tr\,{ab}}$ and the time derivative of $\lambda_{ab}$) actually conspired to cancel one another. But this earlier calculation only succeeded to give an expression for the time derivative of the shear at Scri and did not yield a formula for either $\pi^{\tr\,{ab}}_{~ ~ ~ ,t}$ or $(\mu_\gamma (h^{ac}h^{bd} - \frac{1}{2} h^{ab}h^{cd})\lambda_{cd})_{,t}$ separately at the conformal boundary. To obtain the latter we require some new input.
To derive an expression for $\pi^{\tr\,{ab}}_{~ ~ ~ ,r}$ at Scri, and thereby to complete the derivation of a formula for $\pi^{\tr\,{ab}}_{~ ~ ~ ,t}$ at this boundary, we equate the two expressions for the angular components of the singular terms at Scri (given by Eq. (3.27) and by the angular components of Eq. (5.3)), assuming that they must agree, at least weakly, for any regular solution of the field equations. The resulting formula is:

\newpage

$$\align
&\bigl\{ \frac{\mu_\gamma}{n} \biggl[\bigl(\frac{\pi^{\tr\,{ef}}}{\mu_\gamma}\bigr)_{,r} - \Cal L_{{(2)}_{\tilde Y}} \bigl(\frac{\pi^{\tr\,{ef}}}{\mu_\gamma}\bigr)\biggr] \tag5.10 \\
&- n(Y^eh^{fa} + Y^fh^{ea}) \biggl[\mu_\gamma\frac{(\lambda_{ab}-\frac{1}{2}h_{ab}(h^{cd}\lambda_{cd}))}{n^2}\biggr]^{\vert b} \\
&+ \mu_\gamma \frac{Y^eY^f}{n^2} (\lambda_{ab}\lambda^{ab} -\frac{1}{2} (h^{cd}\lambda_{cd})^2) +\frac{1}{2} \mu_\gamma h^{ef} (\lambda_{ab}\lambda^{ab} -\frac{1}{2} (h^{cd}\lambda_{cd})^2) \\
&+ \frac{3}{2}~ ~  \mu_\gamma (\lambda^{ef} - \frac{1}{2} h^{ef} (h^{cd} \lambda_{cd}))(h^{ab}\lambda_{ab})\bigr\}\bigm|_{\Cal I^{+}} \\
& \overset{\wedge}\to {=} 0
\endalign
$$
where the (weak) vanishing of $\pi^{\tr\,{ri}}\bigm|_{\Cal I^{+}}$ and $h_{ab} \pi^{\tr\,{ab}}\bigm|_{\Cal I^{+}}$ has been imposed to simplify the expression.  

An alternative derivation of the above formula can be developed by considering the conformal transformation properties of the ADM expression (5.5) for the {\it magnetic} components of the Weyl tensor in the analogous way.  Reexpressing this formula in terms of conformal variables, one arrives at
$$\align
& \Omega ~ \gamma^{ms} ~ \Cal B_m^{~ ~ ~ l} \tag5.11 \\
& = \varepsilon^{ijl} ~ \gamma_{ir} ~ \tilde\nabla_j (\gamma) \bigl( \frac{\pi^{\tr\,{rs}}}{\mu_\gamma}\bigr) \\
&- \varepsilon ^{isl} ~ \gamma_{ij} \bigl( \frac{\pi^{\tr\,{pj}}}{\mu_\gamma}\bigr) ~ \frac{\Omega_{,p}}{\Omega} \\
\endalign
$$
The antisymmetric projection, $\varepsilon_{ksl} (\gamma^{ms} \Omega \Cal B_m^{~ ~ ~ l})$, of the above expression vanishes by virtue of its equivalence to the momentum constraint so one needs only to consider the symmetric projection $V^{sl}$ defined by
$$\align
2V^{sl} &:= \gamma^{ms}  ~ \Omega ~ \Cal B_m^{~ ~ ~ l} + \gamma^{ml} \Omega ~ \Cal B_m^{~ ~ ~ s} \tag5.12 \\
&= \varepsilon^{ijl} ~ \gamma_{ik} ~ \tilde\nabla_j (\gamma)  \bigl( \frac{\pi^{\tr\,{ks}}}{\mu_\gamma}\bigr) \\
& + \varepsilon^{ijs} ~ \gamma_{ik} ~ \tilde\nabla_j (\gamma) \bigl( \frac{\pi^{\tr\,{kl}}}{\mu_\gamma}\bigr) .
\endalign
$$
Evaluating the right hand side of this formula and appealing to the regularity constraints $\pi^{\tr\,{ri}}\overset{\wedge}\to{=} 0$ and $[\pi^{\tr\,{ab}} + \mu_\gamma (h^{ac} h^{bd} - \frac{1}{2} h^{ab} h^{cd}) \lambda_{cd}] \overset{\wedge}\to{=} 0$ as well as to Eq. (5.10) and the momentum constraints for the computations of $\pi^{\tr\,{ir}}_{,r}\mid_{\Cal I^{+}}$, one can show by a straightforward calculation that
$$
V^{sl}\bigm|_{r=r_{+}} \overset{\wedge}\to{=} 0 \tag5.13
$$
as would be expected from the assumption that $\Omega \Cal B_l^{~ ~ ~ m}$ vanishes at Scri.
In other words, the vanishing of $\Omega \Cal B_l^{~ ~ ~ m}$ at Scri
necessitates the enforcement of Eq. (5.10) at this boundary.

Finally though, we can remove the guesswork in the aforementioned plausibility argument by appealing to Penrose's well-known result that states that the unphysical Weyl tensor, $\tilde W^\alpha_{~ ~ ~\beta\gamma\delta} (^{(4)}\gamma)$, should vanish at Scri \cite{2,3}. Translated into our notation this result gives directly that $\Omega ~ \Cal E_m^{~ ~ ~ l}$ and  $\Omega ~ \Cal B_m^{~ ~ ~ l}$ vanish there.  To see this, recall that by conformal invariance,
$$\align
W^\alpha_{~ ~ ~ \beta\gamma\delta} (^{(4)}g) &= W^\alpha_{~ ~ ~ \beta\gamma\delta} \bigl( \frac{^{(4)}\gamma}{\Omega^2}\bigr) \tag5.14 \\
& = \tilde W^\alpha_{~ ~ ~ \beta\gamma\delta} (^{(4)} \gamma)
\endalign
$$
and thus that
$$\align
&W_{\alpha\beta\gamma\delta} (^{(4)}g) = ~ {^{(4)}g}_{\alpha\mu} ~ W^\mu_{~ ~ ~ \beta\gamma\delta} (^{(4)}g) \tag5.15 \\
& = \frac{1}{\Omega^2} ~ ^{(4)}\gamma_{\alpha\mu} ~ \tilde W^\mu_{~ ~ ~ \beta\gamma\delta} (^{(4)} \gamma) \\
&= \frac{1}{\Omega^2} ~ \tilde W_{\alpha\beta\gamma\delta} (^{(4)}\gamma).
\endalign
$$
Letting $n^\mu \frac{\partial}{\partial x^{\mu}}$ be the (physical) timelike unit normal field to the chosen slicing (so that $^{(4)}g_{\mu\nu} n^\mu n^\nu = -1$) we get
$$\align
&W_{\alpha\beta\gamma\delta} (^{(4)}g) n^\beta n^\delta \tag5.16 \\
& = \tilde W_{\alpha\beta\gamma\delta} (^{(4)}\gamma) ~ \tilde n^\beta ~ \tilde  n^\delta
\endalign
$$
where $\tilde n^\mu = \frac{n^\mu}{\Omega}$ yields the corresponding unit normal field defined relative to the conformal metric (so that $^{(4)}\gamma_{\mu\nu} \tilde n^\mu ~ \tilde n^\nu = -1$).

The electric components, $E_{ij}$, of the physical Weyl tensor are given (in coordinates for which the chosen slices coincide with $x^0 = t =$ constant hypersurfaces) by
$$\align
E_{ij} &= W_{i\beta j\delta} (^{(4)}g) ~ n^\beta ~ n^\delta \tag5.17 \\
& = \tilde W_{i\beta j\delta} (^{(4)}\gamma)~ \tilde n^\beta ~ \tilde n^\delta
\endalign
$$
and thus vanish at Scri by the aforementioned argument.  Since, however, 
$$\align
\Omega \Cal E_m^{~ ~ ~ l} &= \Omega ~ \mu_g g^{jl} E_{mj} \tag5.18 \\
&= \mu_\gamma ~ \gamma^{jl} E_{mj} 
\endalign
$$
it thus follows that $\Omega \Cal E_m^{~ ~ ~ l}$ vanishes at Scri.  By taking appropriate duals of the above expressions it follows in the same way that $\Omega ~ \Cal B_m^{~ ~ ~ l}$ also vanishes there.  Either one of these results, as we have seen, is sufficient to imply Eq. (5.10) for the radial derivative of $\pi^{\tr\,{ab}}$ at the outer boundary.

Though somewhat peripheral to the above discussion, we conclude this section with some remarks on the evolution of geometric data along the conformal boundary and on the choice of a ``conformal gauge'' at Scri.  Recalling that the shift vector components $X^i$ satisfy (c.f., Eqs. (2.18), (3.9), (7.4))
$$\align
X^r\bigm|_{r=r_{+}} &\overset{\wedge}\to{=}  ~ - ~ \frac{\tilde N}{n}\biggm|_{r=r_{+}} \tag5.19 \\
X^d\bigm|_{r=r_{+}} &\overset{\wedge}\to{=} \left\{ \frac{\tilde N}{n} ~ Y^d ~ + Z^d\right\}\biggm|_{r=r_{+}} 
\endalign
$$
one sees that
$$
(X^r ~ Y^c + X^c)\biggm|_{r=r_{+}} \overset{\wedge}\to{=} Z^c\biggm|_{r=r_{+}}. \tag5.20
$$
Using this, it follows from a direct evaluation of the angular component of $\gamma_{ij,t}$ at Scri that the metric $h_{ab}$ induced on $2$-dimensional slices of Scri satisfies the evolution equation
$$\align
& h_{ab,t}\biggm|_{r=r_{+}} \overset{\wedge}\to{=} \{ \frac{2\tilde N}{\mu_\gamma} ~ h_{ac} ~ h_{bd} [ \pi^{\tr\,{cd}} \tag5.21 \\
& + \mu_\gamma (h^{ce} h^{df} - ~ \frac{1}{2} ~ h^{cd} ~ h^{ef})\lambda_{ef} ] \\
& + \frac{1}{3} ~ \tilde N h_{ab} ( \frac{\Gamma}{\tilde N} ~ + 3 ~ h^{ef} \lambda_{ef}) \\
& + \frac{2\tilde N}{\mu_\gamma} ( Y_aY_b \pi^{\tr\,{rr}} + Y_a h_{bd} \pi^{\tr\,{rd}} + Y_b h_{ad} \pi^{\tr\,{dr}}) \\
& + Z^c h_{ab,c} + Z^c_{,a} h_{cb} + Z^c_{,b} h_{ac}\}\biggm|_{r=r_{+}}
\endalign
$$
which clearly is weakly equivalent to
$$\align
h_{ab,t}\biggm|_{r=r_{+}} &\overset{\wedge}\to{=} \{ \frac{\tilde N}{3} ~ h_{ab} (\frac{\Gamma}{\tilde N} + 3(h^{ef} \lambda_{ef})) \tag5.22 \\
&+ \Cal L_{{(2)}_{\tilde Z}} ~ h_{ab} \}\biggm|_{r=r_{+}}.
\endalign
$$
One could thus exploit the freedom to choose Dirichlet data for the function $\Gamma$ (which heretofore has remained unconstrained) to arrange that $(\Gamma + 3 \tilde N ~ h^{ef} ~ \lambda_{ef})\bigm|_{r=r_{+}}$ vanishes on each $t = $ constant slice of Scri.  This choice would reduce the above evolution equation to the essentially trivial form
$$
h_{ab,t} \biggm|_{r=r_{+}} \overset{\wedge}\to{=} \{ \Cal L_{{(2)}_{\tilde Z}} ~ h_{ab} \} \biggm|_{r=r_{+}} \tag5.23
$$
or, if one also exploits the freedom to set $Z^d\biggm|_{r=r_{+}} \overset{\wedge}\to{=} 0$ at Scri, to the manifestly trivial form
$$
h_{ab,t}\biggm|_{r=r_{+}} \overset{\wedge}\to{=} 0 \tag5.24
$$
when $Z^d \biggm|_{r=r_{+}} = (\Gamma + 3\tilde N ~ h^{ef} \lambda_{ef})\biggm|_{r=r_{+}} \overset{\wedge}\to{=} 0$.

A `covariant interpretation' of this (potential) choice for $(\Gamma + 3\tilde N ~ h^{ef} \lambda_{ef})\biggm|_{r=r_{+}}$ can be uncovered by applying the (conformal) wave operator $\square_{(4)_\gamma}$ to the conformal factor $\Omega$ and evaluating the result at Scri.  A straightforward calculation yields the result
$$
\square_{(4)_\gamma} \Omega\biggm|_{\Cal I^{+}} \overset{\wedge}\to{=} ~ \frac{2K}{3} \left\{ h^{ab} \lambda_{ab} + \frac{1}{3} ~ \frac{\Gamma}{\tilde N} \right\} \biggm|_{r=r_{+}}. \tag5.25
$$
Thus the special choice of Dirichlet data for $\Gamma$ discussed above corresponds to imposing the `conformal gauge condition' that $\square_{(4)_\gamma} ~\Omega$ vanish at Scri.  By appealing to the conformal transformation properties of the $4$-dimensional Einstein tensor, one can show that this conformal gauge choice further implies that
$$
^{(4)}\tilde\nabla_\mu (^{(4)}\gamma) ~ ^{(4)}\tilde\nabla_\nu ~ (^{(4)}\gamma) \Omega\biggm|_{\Cal I^{+}} \overset{\wedge}\to{=} 0\tag5.26
$$
i.e., that the full $4$-dimensional conformal Hessian of $\Omega$ vanishes at Scri \cite{16}.  Since this result simplifies numerous formulas evaluated at Scri, it is often imposed to streamline various calculations at the conformal boundary.  The above discussion shows that such a choice of conformal gauge is fully compatible with our formalism and determines a correspondingly `natural' choice for the Dirichlet data for $\Gamma$ while leaving unconstrained the boundary values of $\tilde N$ and $Z^d$. 

Assuming Penrose's result that the quantities $\Omega\Cal E_m^{~ ~ ~ l}$ and $\Omega\Cal B_m^{~ ~ ~ l}$ should vanish at Scri we can use Eq. (5.10) to eliminate the radial derivative, $\pi^{\tr\,{ef}}_{,r}\bigm|_{\Cal I^+}$, from the equation of motion for $\pi^{\tr\,{ef}}\bigm|_{\Cal I^+}$.  In view of Eqs. (4.6) and (5.21) however, it clearly suffices to present instead the (slightly simpler) equation of motion for $\lambda_{ab}\bigm|_{\Cal I^+}$.  A straightforward calculation making use of Eqs. (2.21), (3.2) and (3.7) yields
$$\align
&\{ 2n\frac{\partial\lambda_{ab}}{\partial t}\} \bigm|_{r=r_+} \overset{\wedge}\to{=} \tag5.27 \\
&\{ -n\lambda_{ab} \frac{2\tilde N}{\mu_\gamma} n^2 \pi^{\tr\,{rr}} + \frac{n}{3} \Gamma \lambda_{ab} \\
&+ [ 2Y_b \frac{\tilde N}{\mu_\gamma} n^2 \pi^{\tr\,{rr}} + \frac{2\tilde N}{\mu_\gamma} n^2 h_{cb} \pi^{\tr\,{rc}}]_{\vert a} \\
&+ [2Y_a \frac{\tilde N}{\mu_\gamma} n^2 \pi^{\tr\,{rr}} + \frac{2\tilde N}{\mu_\gamma} n^2 h_{ac} \pi^{\tr\,{rc}} ]_{\vert b} \\
&- \{ \frac{2\tilde N}{\mu_\gamma} h_{ac} h_{bd} \pi^{\tr\,{cd}} + \frac{2\tilde N}{\mu_\gamma} [Y_aY_b \pi^{\tr\,{rr}} \\
&+ Y_a h_{bd} \pi^{\tr\,{rd}} + Y_b h_{ad} \pi^{\tr\,{rd}}]\}_{,r} \\
&- 2\tilde N \lambda_{ab,r} + 2\tilde N (\Cal L_{{}^{(2)} \tilde Y}\lambda_{ab}) + 2n(\Cal L_{{}^{(2)} \tilde Z}\lambda_{ab})\\
&+ \Cal L_{{}^{(2)} \tilde Y} [\frac{2\tilde N}{\mu_\gamma} h_{ac} h_{bd} \pi^{\tr\,{cd}} + \frac{2\tilde N}{\mu_\gamma}(Y_aY_b \pi^{\tr\,{rr}} \\
&+ Y_a h_{bd} \pi^{\tr\,{rd}} + Y_b h_{ad} \pi^{\tr\,{rd}})] \\
&- [n^2 (\frac{\tilde N}{n})_{\vert b}]_{\vert a} - [n^2 (\frac{\tilde N}{n})_{\vert a}]_{\vert b} \\
&+ \frac{1}{3} h_{ab} (Y^c\Gamma_{,c} - \Gamma_{,r})\}\bigm|_{r=r_+}.
\endalign
$$
For completeness we also list the equations of motion for the metric functions $n$ and $Y^c$:
$$\align
&\qquad \frac{2}{n} \frac{\partial n}{\partial t} = \frac{2\tilde N}{\mu_\gamma} n^2 ~ \pi^{\tr\,{rr}} + \frac{1}{3} 
\Gamma \tag5.28 \\
&+ 2 X^c \frac{n_{,c}}{n} - 2 Y^c X^r_{,c} + \frac{2}{n} (n X^r)_{,r} ~ ~, \\
\endalign
$$
$$\align
&\frac{\partial Y^c}{\partial t} = 2 Y^c \frac{\tilde N}{\mu_\gamma} n^2 \pi^{\tr\,{rr}} + \frac{2\tilde N}{\mu_\gamma} n^2 \pi^{\tr\,{rc}} \tag5.29 \\
&+ X^a Y^c_{,a} - X^c_{,a} Y^a + X^r_{,a} (n^2 h^{ac} - Y^aY^c)    \\
&+ (X^r Y^c + X^c)_{,r}.
\endalign
$$
Notice though that even when these are evaluated at $\Cal I^+$ they involve radial derivatives of the unknowns as well as those of $X^i$ at Scri.  The same is true of Eq. (5.27).  Thus, unlike Eqs. (4.1) or (5.21) for example, Eqs. (5.27) - (5.29), when evaluated at Scri, are not fully intrinsic to this boundary.  This would of course be all the more true if Penrose's result is not presupposed and the radial derivatives, $\pi^{\tr\,{ef}}_{,r}\bigm|_{\Cal I^+}$, can therefore not be eliminated as discussed above.  On the other hand, if Penrose's smoothness assumptions are made, then one can return to the momentum constraints (3.17) and compute an additional radial derivative thereof in order to derive expressions for $\pi^{\tr\,{rr}}_{,rr}\bigm|_{\Cal I^+}$ and $\pi^{\tr\,{ra}}_{,rr}\bigm|_{\Cal I^+}$.  These may be useful in extending the Taylor expansions for the quantities $\pi^{\tr\,{ri}}$ about $\Cal I^+$ but one should keep in mind that the additional smoothness requirements needed to justify this calculation may be incompatible with the generality desired for the solutions under study.  We shall return briefly to this issue in the concluding section.

\head VI. Concluding Remarks\endhead
\vskip .10in

In the section above we mentioned the (still controversial) issue of deciding whether Penrose's smoothness hypotheses are overly restrictive from the stand-point of physical generality.  Fortunately however most of our central results do not require stringent differentiability assumptions and are in fact compatible with so-called `polyhomogeneous' (or `polylogarithmic') behavior at Scri.  To see this we appeal to the conclusions derived in Refs. \cite{12,13} wherein one finds that polylogarithmic behavior, if present, sets in at the level of the fourth radial derivative of $\Omega$ and the second radial derivative of $\pi^{\tr\,{ij}}$ at $\Cal I^+$.  But our Taylor expansion calculations only determined the radial derivatives of $\Omega$ up to the third order and those of $\pi^{\tr\,{ri}}$ up to the first.  Only if Penrose-level regularity was additionally assumed could one go further and compute the quantities $\{ \pi^{\tr\,{rr}}_{,rr}, \pi^{\tr\,{ra}}_{, rr}, \pi^{\tr\,{ab}}_{,r}\}\bigm|_{\Cal I^{+}}$ for example, but none of these were needed in the demonstration of regularity at the boundary or its preservation under time evolution.  

If polylogarithmic Scris are allowed, then one must forego the argument that the unphysical Weyl tensor vanishes at null infinity and therefore also the corresponding evaluation of the aforementioned, higher radial derivatives at $\Cal I^+$ since the latter, in general, may not have well-defined limits at this boundary.  But since none of these were essential to our central results, we believe the latter are fully compatible with polylogarithmic, as opposed to smooth, behavior at Scri.

Some numerical relativists may wish to put their outer boundaries at Scri but prefer not to adopt constrained evolution or prefer to avoid the use of our CMCSH gauge conditions and wonder, accordingly, whether our results have any relevance for them. While we cannot draw definitive conclusions that are sure to apply to an arbitrary numerical setup we can nevertheless make several general points in this regard. First of all, the main tool needed for verifying finiteness and then actually evaluating the apparently singular terms at Scri has been the strict enforcement, at least to the requisite order in a Taylor expansion, of the Einstein constraint equations and the vanishing shear condition at the outer boundary. Thus we would anticipate that any strategy for say free (as opposed to constrained) evolution would have to be coupled with strict enforcement, to the needed order, of these constraints at the boundary. Otherwise, there seems to be no hope for showing that the evolution equations are actually regular at Scri. On the other hand, the `universal' character of our calculations (i.e., the fact that they hinged purely on imposition of these constraints) would seem to show that they apply equally well to a variety of formulations of the field equations and not just our own.

Furthermore, gauge conditions played a rather peripheral role in our analysis and, for that reason, we do not think that our particular choice was at all crucial for the main results. We have already sketched how the CMC condition could be relaxed and our calculations modified accordingly. If the mean curvature, $\tr_gK = g_{ij} K^{ij}$, is allowed to evolve, its equation of motion, in our notation, can be written
$$\align
&\frac{\partial}{\partial t} (\tr_g K) = -\Omega (\gamma^{ij} \tilde N_{;ij}) - \tilde N(\gamma^{ij} \Omega_{;ij}) - 
\frac{\tilde R(\gamma)}{2} \Omega \tilde N \tag6.1 \\
&\qquad + 3\gamma^{ij} \tilde N_{,i} \Omega_{,j}  + \frac{3}{2} \tilde N \Omega 
\frac{\pi^{\tr\,{im}}\pi^{\tr\,{jl}}}{(\mu_\gamma)^2} ~ \gamma_{ij} \gamma_{ml}\\
&\qquad + X^i (\tr_g K)_{,i}
\endalign
$$
and has no irregular behavior at Scri. Furthermore, our preservation of regularity calculations did not make use of our elliptic equation for the shift vector $X^i \frac{\partial}{\partial x^i}$ (imposed to preserve spatial harmonic coordinate conditions for $\gamma_{ij}$) or of our normalization condition $\tilde R(\gamma) =$ constant. In addition, they did not lead to any restriction upon the asymptotic gauge data $\{ \tilde N, Z^a \frac{\partial}{\partial x^a}\}\bigm|_{\Cal I^{+}}$. They did make use of our elliptic  equation for the conformal lapse function $\tilde N$ but this was forced in large measure by our imposition of CMC slicing conditions and use of a simplified form of the momentum constraint which only holds in CMC gauge. We have little doubt that the preservation of regularity calculations would work equally well in a variety of other frameworks.

Concerning the issue of a stable, numerical implementation of the regularity conditions at Scri, this has not yet been carried out. A key point though is that, in any such implementation, the computer would {\it not} be expected to re-derive (by delicate limiting procedures) regular expressions for the apparently singular terms in the evolution equations at the boundary.  These would be explicitly provided as boundary conditions through formulas such as (3.25)-(3.27). In view of the comments in section IV regarding the 'frictional restoring forces' in effect near Scri, it is very plausible that a stable implementation of these boundary conditions is numerically achievable.

In summary, we see little reason why a wide variety of numerical setups could not consistently place their outer boundaries at Scri. As a prominent numerical relativist recommended to one of us some years ago ``If you want to go to Scri - just do it''.
\vskip .15in

\head VII. Appendix\endhead
\vskip .10in
 
In the notation we have used, the Einstein tensor,
$$
\tilde G^{ij} (\gamma) := \tilde R^{ij} (\gamma) - \frac{1}{2} \gamma^{ij} \tilde R (\gamma), \tag7.1
$$
of the metric $\gamma_{ij}$ has components
$$\align
&\tilde G^{rr}(\gamma) = - ~ ~ \frac{1}{2n^2} \bigl[\lambda_{ab}\lambda_{cd}h^{ac}h^{bd} - (h^{cd}\lambda_{cd})^2 \\
&\qquad\qquad\qquad\qquad\quad + \; {^{(2)} R}(h)\bigr], \tag7.2 \\
&\tilde G^{rf}(\gamma) = \frac{1}{n} h^{af} \bigl[ - ~ ~ \lambda_{ab}^{~ ~ ~\vert b} + (h^{cd} \lambda_{cd})_{\vert a}\bigr] \\
&\qquad + \frac{Y^f}{2n^2} \bigr[\lambda_{ab}\lambda_{cd}h^{ac}h^{bd} - (h^{cd}\lambda_{cd})^2 + ~ ^{(2)}R(h)\bigr],\\
&\tilde G^{ef}(\gamma) = \biggl(\frac{h^{ea}h^{fb}}{n} -
\frac{h^{ef}h^{ab}}{n}\biggr) [\lambda_{ab,r} - \Cal L_{{}^{(2)} \tilde Y} \lambda_{ab} \\
&- n_{\mid ab} + n ~ ^{(2)}R_{ab}(h) + 2nh^{cd} \lambda_{ad}\lambda_{bc} - n\lambda_{ab} (h^{cd}\lambda_{cd})\bigr]\\
& - \frac{1}{n} (Y^eh^{af} + Y^f h^{ea})(-\lambda_{ac}^{~ ~ ~ \vert c} + (h^{cd} \lambda_{cd})_{\vert a}) \\
& + \frac{1}{2} \bigl( h^{ef} - \frac{Y^eY^f}{n^2}\bigr)\bigl[\lambda_{ab}\lambda_{cd} h^{ac}h^{bd} - (h^{cd}\lambda_{cd})^2 \\
&\qquad\qquad\qquad\qquad\quad + ~ ^{(2)}R(h)\bigr]
\endalign
$$
where $^{(2)}R_{ab}(h)$ and $^{(2)}R(h)$ are the Ricci tensor and
scalar curvature of the $2$-metric $h_{ab}$, $\vert a$ signifies
covariant differentiation with respect to this metric and $\Cal
L_{{}^{(2)} \tilde Y}$ signifies Lie differentiation with respect to
the $2$-dimensional `shift' field $Y^a \frac{\partial}{\partial x^a} =
{}^{(2)} \tilde Y$. The Laplacian of a function $F$ is given by
$$\align
&\gamma^{ij} F_{;ij} = (h^{ab} + \frac{Y^aY^b}{n^2}) ~ F_{\vert ab} - \frac{F_{,r}}{n} (h^{ab}\lambda_{ab}) \tag7.3 \\
& +\frac{Y^c}{n} F_{,c} (h^{ab}\lambda_{ab}) + F_{,r} ~ \frac{Y^a}{n^3} n_{\mid a} - F_{,r} ~ \frac{n_{,r}}{n^3} \\
& - \frac{2}{n} Y^a F_{,c} h^{cb} \lambda_{ab} + \frac{2Y^a}{n^2} F_{,c} Y^c_{~ ~ ~ \vert a} \\
& + \frac{F_{,c}}{n} h^{cd} n_{\vert d} - \frac{Y^a}{n^3} F_{,c} Y^c n_{\vert a} - \frac{2Y^a}{n^2} F_{,ra} \\
& + \frac{1}{n^2} F_{,rr} + \frac{F_{,c}}{n^3} Y^c n_{,r} - \frac{F_{,c}}{n^2} h^{cd} Y_{d,r} \\
& + \frac{F_{,c}}{2n^2} (Y^a Y^b h_{ab})^{\vert c} .
\endalign
$$
The contravariant components, $\gamma^{ij}$, of this metric are given by
$$\align
&\gamma^{rr} = \frac{1}{n^2} , ~ \gamma^{ra} = - ~ ~ \frac{Y^a}{n^2},\tag7.4 \\
& \gamma^{ab} = h^{ab} + \frac{Y^aY^b}{n^2} 
\endalign
$$
(where $h^{ab}$ is the contravariant form of $h_{ab}$) and the volume element $\mu_\gamma$ is expressible as
$$
\mu_\gamma = n\mu_h \tag7.5 
$$
where $\mu_h$ is the area element of $h_{ab}$.

\medskip

\centerline{\bf Acknowledgements}
\medskip

The authors are grateful to Lars Andersson, Marcus Ansorg, Robert
Beig, Luisa Buchman, Piotr Chru\' sciel, Thibault Damour, Helmut
Friedrich, J\"org Hennig, Michael Holst,  Harald Pfeiffer, Olivier
Sarbach, Nikodem Szpak, Jeffrey Winicour and Anil Zengino\u glu for
numerous, valuable comments and extensive helpful discussions.
Vincent Moncrief is particularly grateful to the Albert Einstein
Institute in Golm, Germany, the Universit\"at Wien in Vienna, Austria,
the Institut des Hautes \'Etudes Scientifiques in Bures-sur-Yvette,
France, the Mittag-Leffler Institute in Djursholm, Sweden and the California Institute of Technology in Pasadena, California for the hospitality and support given to him during visits where portions of this research were carried out. Oliver Rinne gratefully acknowledges funding through a Research Fellowship at King's College, Cambridge. This research was also supported by the National Science Foundation through grants PHY-0354391 and PHY-0647331 to Yale University. Rinne's research was also supported by NSF grant PHY-0601459 and NASA grant NNG05GG52G and a grant from the Sherman Fairchild Foundation to Caltech.

\end
\newpage


\Refs

\ref \no 1 
\by Penrose, R. 
\paper Conformal Treatment of Infinity
\inbook Relativity, Groups and Topology
\eds C. DeWitt and B. DeWitt
\publ Gordon and Breach
\publaddr New York
\yr 1964
\pages 565--584
\endref
\vskip .10in

\ref \no 2 
\bysame 
\pages 159--203
\paper Zero rest-mass fields including gravitation: asymptotic behavior
\jour Proc. R. Soc. London Ser. A
\vol 284
\yr 1965
\endref
\vskip .10in

\ref \no 3 
\by Geroch, R. 
\pages 1--105
\paper Asymptotic Structure of Space-time
\inbook Asymptotic Structure of Space Time
\eds Esposito, F.P., and Witten, L. 
\publ Plenum Press
\publaddr New York
\yr 1977
\endref
\vskip .10in

\ref \no 4 
\by Bondi, H., van der Burg, M.G.J., and Metzner, A.W.K.  
\pages 21--52
\paper Gravitational waves in general relativity 
  VII. Waves from axi-symmetric isolated systems
\jour Proc. R. Soc. London Ser. A
\vol 269 
\yr 1962
\endref
\vskip .10in

\ref \no 5 
\by Friedrich, H.
\paper Cauchy problems for the conformal vacuum field equations in 
  general relativity
\jour Commun. Math. Phys.
\vol 91
\pages 445-472
\yr 1983
\moreref 
\paper On the Regular and the Asymptotic Characteristic Initial Value 
  Problem for Einstein's Vacuum Field Equations
\jour Proc. R. Soc. London Ser. A
\vol 375
\pages 169--184 
\yr 1981
\endref
\vskip .10in

\ref \no 6 
\by For a comprehensive review of these results including many 
  of the relevant references, see 
  Frauendiener, J.
\paper Conformal Infinity 
\jour Living Rev. Relativity
\vol 7
\yr 2004
\finalinfo 1. http://www.livingreviews.org/\ lrr-2004-1
\endref
\vskip .10in

\ref \no 7 
\by Andersson, L. and Moncrief, V. 
\pages 1--34
\paper Elliptic Hyperbolic Systems and the Einstein  Equations
\jour Ann. Henri Poincar\'e 
\vol 4 
\yr 2003
\endref
\vskip .10in

\ref \no 8
\by Rinne, O.
\paper Constrained evolution in axisymmetry and the gravitational collapse 
  of prolate Brill waves
\jour Class. Quantum Grav. 
\vol 25
\page 135009 
\yr 2008
\endref
\vskip .10in

\ref \no 9
\by Buchman, L. and Pfeiffer, H.
\paperinfo in preparation
\endref
\vskip .10in

\ref \no 10
\by Moncrief, V.
\paperinfo in preparation 
\endref
\vskip .10in

\ref \no 11
\by Escobar, J.F.
\paper Conformal deformation of a Riemannian metric to a scalar
  flat metric with constant mean curvature on the boundary
\jour Ann. of Math. (2) 
\vol 136
\pages 1--50 
\yr 1992
\moreref { }
\by Escobar, J.F. and Garcia, G.
\paper Conformal metrics on the ball with zero scalar curvature and 
  prescribed mean curvature on the boundary
\jour J. Funct. Anal. 
\vol 211
\pages 71--152 
\yr 2004
\endref
\vskip .10in

\ref \no 12
\by Andersson, L., Chru\'sciel, P.T. and Friedrich, H. 
\pages 587--612
\paper On the regularity of solutions to the Yamabe equation and the 
  existence of smooth hyperboloidal initial data for Einstein's field equations
\jour Commun. Math. Phys.
\vol 149
\yr 1992
\endref
\vskip .10in

\ref \no 13
\by Andersson, L., and Chru\'sciel, P.T.
\paper On hyperboloidal Cauchy data for vacuum Einstein equations 
  and obstructions to smoothness of scri
\jour Commun. Math. Phys. 
\vol 161
\pages 533--568
\yr 1994
\endref
\vskip .10in

\ref \no 14
\by  Christodoulou, D. and Klainerman, S.
\book The global non-linear stability of Minkowski space
\publ Princeton University Press
\publaddr Princeton
\yr 1993
\endref
\vskip .10in

\ref \no 15
\by Andersson, L. and Moncrief, V.
\paper Future Complete Vacuum Spacetimes
\inbook The Einstein Equations and the Large Scale Behavior of 
  Gravitational Fields 
\eds P.T. Chru\'sciel and H. Friedrich 
\publ Birkh\"auser
\publaddr Basel 
\pages 299-330 
\yr 2004
\endref
\vskip .10in

\ref \no 16 \by Zengino\u glu, A.
\paper Hyperboloidal evolution with the Einstein equations
\jour Class. Quantum Grav. 
\vol 25
\yr 2008
\page 195025
\endref
